\documentclass[11pt,a4paper]{article}
\pdfoutput=1
\usepackage{jcappub}
\usepackage{ifthen}
\usepackage[T1]{fontenc}
\usepackage{lmodern}
\usepackage[utf8]{inputenc}
\usepackage[english]{babel}
\usepackage[usenames,dvipsnames,svgnames]{xcolor}
\usepackage{amsmath,amssymb}
\usepackage{array,multirow,graphicx}
\graphicspath{{figures/}}
\usepackage{float}
\usepackage{subfig}
\usepackage{comment}

\usepackage{cancel}
\usepackage[colorinlistoftodos]{todonotes}
\usepackage{import}

\usepackage{natbib}

\newcommand{\thetac}{\theta_T}
\newcommand{\thetas}{\theta_\ast}
\newcommand{\eps}{\varepsilon}

\title {\huge The CMB Cold Spot under the lens II: \\
{\huge Lensing signatures in polarization and cosmic texture footprints}}

\author[a,b,c,d]{Pedro da Silveira Ferreira,}
\author[e]{Stephen Owusu}

\affiliation[a]{Center for Cosmology and Computational Astrophysics, Institute for Advanced Study in Physics, Zhejiang University, Hangzhou 310058, China.}
\affiliation[b]{Institute of Astronomy, School of Physics, Zhejiang University, Hangzhou 310058, China}
\affiliation[c]{Observat\'orio do Valongo, Universidade Federal do Rio de Janeiro, 20080-090, Rio de Janeiro, RJ, Brazil}
\affiliation[d]{Centro Brasileiro de Pesquisas Físicas, 22290-180, Rio de Janeiro, RJ, Brazil}
\affiliation[e]{North Hennepin Community College, 55445,  Brooklyn Park, Minnesota, United States of America}

\emailAdd{dasferreira.pedro@gmail.com}
\emailAdd{stephen.owusu@nhcc.edu}

\abstract{
We forecast the detectability of the lensing footprint of a collapsing cosmic texture, a topological defect proposed as an explanation of the CMB Cold Spot. Our pipeline is a quadratic, template-amplitude estimator for localized, azimuthally symmetric lensing profiles: it projects the standard off-diagonal covariance response of lensed CMB fields onto a physically motivated template. Rather than reconstructing an arbitrary lensing field, the method targets weak but coherent localized footprints from sources such as voids, clusters and topological defects. Using the temperature/polarization pairs $TT$, $TE+ET$, $EE$, $TB+BT$ and $EB+BE$ for forthcoming Simons Observatory data, we estimate a $2.3\sigma$ detection if the texture amplitude reaches the current \textit{Planck} 2018 $2\sigma$ upper limit, and a $1.5\sigma$ measurement for the best-fit texture parameters. This sensitivity is notable given the expected typical deflection angle below $6''$. The inclusion of polarization substantially increases the cumulative signal-to-noise ratio, by $\sim67\%$ relative to temperature alone, making sub-$10''$ localized lensing footprints accessible to SO-like surveys.
}


\begin{document}

\notoc
\maketitle

\section{Introduction} \label{sec:intro}

Upcoming ground-based CMB experiments promise a transformative leap beyond \textit{Planck}, delivering arcminute-scale resolution and high-fidelity polarization maps \cite{SimonsObservatory:2025wwn, macinnis2024cosmological}. Among them, the Simons Observatory (SO) deployed a suite of small- and large-aperture telescopes on Cerro Toco in 2025, targeting noise levels of $\Delta_T\simeq2.6\,\mu{\rm K}\cdot\mathrm{arcmin}$ and beam sizes as small as $1\hbox{--}2'$ at 90--280 GHz \cite{SimonsObservatory:2025wwn}. These specifications unlock multipoles up to $\ell\!>\!3000$ in polarization and $\ell\!>\!4000$ in temperature, whereas \textit{Planck} is noise-dominated for $\ell\!\gtrsim\!1000$ in polarization and $\ell\!\gtrsim\!1800$ for temperature \cite{Planck:2019nip}. Crucially, high-S/N CMB temperature and polarization maps will sharpen the sensitivity of lensing estimators to arcsecond-scale deflections, enabling searches for exotic physics that imprints weak but coherent patterns on the microwave sky.

In this work we extend the standard temperature pipeline presented in \cite{Owusu:2022etl} to a template-amplitude estimator that uses temperature, polarization and cross-correlation off-diagonal modes. Rather than reconstructing an arbitrary lensing field, the pipeline projects the quadratic lensing response onto a localized profile. This makes it agnostic to the detailed mass model while remaining optimized for weak, coherent signatures, such as the $\lesssim\!10''$ deflections expected from clusters, voids and, central to this study, collapsing cosmic textures.

\begin{figure}[h!]
    \centering
    \includegraphics[width=0.74\linewidth]{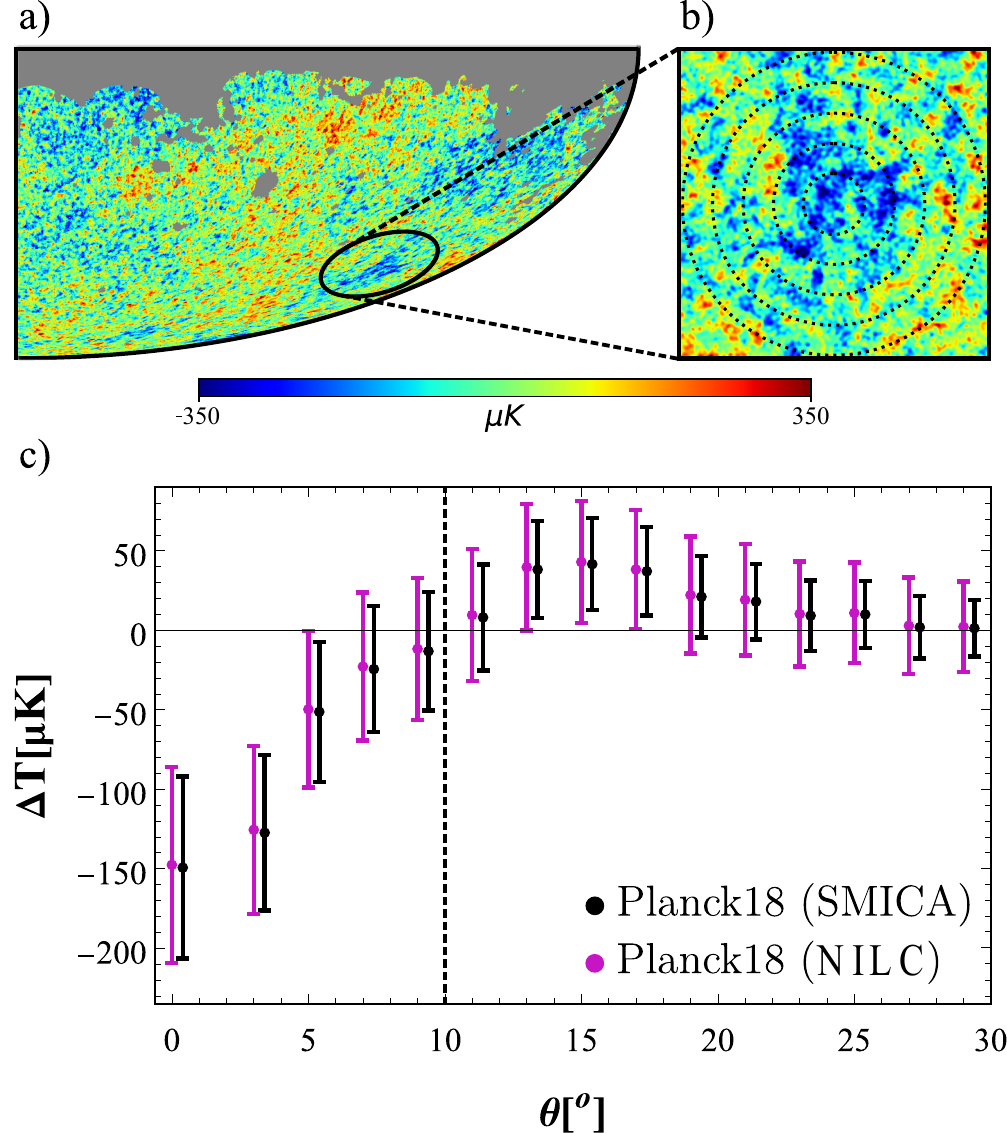}
    \caption{\textbf{a)} Bottom right quadrant of the \textit{Planck} 2018 CMB temperature map (SMICA pipeline \cite{Akrami:2018mcd}) in Mollweide projection. The colors indicate the temperature variations around the mean. The black ellipse represents the region of the CS. \textbf{b)} Zoom of the CS region. The plot spans an area of $10^{\circ}\times10^{\circ}$. The space between dashed circles indicates the rings of $2^\circ$ width used to compute the CS temperature profile. \textbf{c)} The temperature profile obtained by averaging the pixels inside these rings up to $\theta=30^\circ$ for two different component separation pipelines (SMICA and NILC). The dashed line indicates the region shown in panel b). The error bars are the standard deviation of each ring computed using DX12 \textit{Planck} simulations~\cite{PlanckFFP10}.
    }\label{fig:cs_position_and_profile}
\end{figure}

Our science target for the estimator is the Cosmic Microwave Background (CMB) Cold Spot (CS). The CS is a region in the CMB that appears significantly colder than the surrounding areas, deviating from the expected isotropic distribution predicted by the standard model of cosmology. It is located at Galactic coordinates $(l, b)\sim (209^\circ, 57^\circ)$ and extends approximately $10^\circ$ in angular radius, see Figure~\ref{fig:cs_position_and_profile}. Since its initial identification in the WMAP data~\cite{Vielva:2003et}, the anomaly has been the subject of numerous analyses~\cite{Cruz:2004ce,Cruz:2006sv,Cruz:2008sb}. Observations from the \textit{Planck} satellite reaffirmed the Cold Spot and placed its significance at roughly \(3\sigma\) when evaluated against the predictions of the \(\Lambda\)CDM model~\citep{Planck:2015igc}. Unlike random Gaussian fluctuations predicted by the standard cosmological model $\Lambda$CDM, the CS exhibits a coherent temperature profile that is difficult to attribute to random fluctuations~\cite{vielva2010comprehensive, cruz2009wmap}.

A purely Galactic-foreground explanation of the CS is unlikely, as its spectrum remains constant with frequency, and it lies in a region with minimal foreground contamination~\citep{Cruz:2006sv}. However, recently \cite{lambas2024cmb} presented evidence that the CMB Cold Spot could be largely attributable to foreground effects from galaxies, especially those within the Eridanus supergroup. Similarly, the Sunyaev--Zel'dovich effect is an unlikely cause, given the absence of any major low-redshift clusters in that direction.
Several alternative explanations have been proposed, including a cosmic texture or a large void along the line of sight. However, ~\citep{Owusu:2022etl} ruled out the possibility that the CS is due to a large void between us and the surface of last scattering. The evidence ratio favored the standard $\Lambda$CDM model over the supervoid hypothesis, with odds of 1:13 (1:20) for SMICA (NILC) maps \cite{Akrami:2018mcd}, compared to the original odds 56:1 (21:1) using temperature data alone.

Here we focus on the possibility that the CS is caused by a collapsing cosmic texture. In the early universe, at extraordinarily high energies ($\sim10^{15}$--$10^{16}$ GeV), it is hypothesized that all fundamental forces, excluding gravity, were unified under a single framework referred to as the Grand Unified Theory (GUT). This theory is described by a larger symmetry group, such as SU(5), SU(10), among others~\cite{georgi1974unity, vilenkin1994cosmic}. As the universe expands and cools, it undergoes a phase transition leading to the separation of these forces in a process known as spontaneous symmetry breaking~\cite{kibble1976topology, vilenkin1994cosmic}. These transitions can lead to the formation of topological defects, including cosmic textures: unstable structures that can collapse and induce perturbations in the metric of spacetime, potentially seeding large-scale structure~\cite{Turok:1990gw}. These perturbations affect the path of photons, resulting in anisotropies that can appear as characteristic hot or cold spots in the CMB.  

In addition to the temperature shift, the lensing of the CMB photons caused by the gravitational field generated by the collapsing texture has also been studied~\cite{Das:2008es,Farhang:2020gsa}. Such lensing has been shown to introduce correlations in the off-diagonal two-point and three-point functions~\cite{Masina:2009wt, Masina:2010dc}. However, the expected lensing angle for a texture capable of generating the CS is $\sim50$ times less than that produced by a void. Another approach to check for the presence of a cosmic texture is to compare the anisotropy profile in both temperature and polarization maps \cite{Vielva:2010vn}. Given the tenuous footprint of such an effect, any gain in S/N and any independent diagnostic should be considered.  

In anticipation of the improvements from upcoming CMB experiments, which will deliver substantially lower-noise and higher-resolution polarization maps than \textit{Planck}, we introduce a polarization-based quadratic estimator tailored to a \emph{localized, azimuthally symmetric} lensing potential centered on the CMB Cold Spot. Our focus is not to reconstruct the full lensing potential field $\phi_{LM}$. Instead we compress the lensing-induced off-diagonal mode coupling into a small set of physically motivated template parameters (here, the texture parameters), yielding a single, directly interpretable detection statistic and an efficient forecasting pipeline for targeted localized lenses. Because the pipeline only assumes circular symmetry and a specified center, it can also be applied to other localized lensing imprints of known morphology, or generalized to a sky scan to search for candidates.

Conceptually, our statistic is the maximum-likelihood matched-filter amplitude of a fixed lensing template. It can therefore be viewed as the standard Hu--Okamoto CMB lensing quadratic estimator \cite{Okamoto:2003zw} projected onto a known localized profile, closely analogous to matched-filter approaches used in CMB cluster lensing. The new ingredients relative to our previous temperature-only analysis are the projection onto the texture lensing profile with the updated SO specifications and the explicit inclusion of the ordered polarization channels $TE$, $ET$, $EE$, $TB$, $BT$, $EB$ and $BE$. 

Throughout this work we adopt a Hu--Okamoto--type inverse-variance weighting within each ordered pair, rather than the strictly optimal global-minimum-variance (GMV) combination of quadratic estimators~\cite{Maniyar:2021msb}. A full GMV implementation would reduce the variance by using the inverse of the complete covariance matrix among all quadratic pairs. As shown in Ref.~\cite{Maniyar:2021msb}, for SO-like noise levels the improvement in reconstruction noise from Hu--Okamoto to GMV weighting is modest (up to $9-12\%$ depending on scale). Our pair-by-pair weighting is therefore less optimal but transparent: it keeps the contribution of each channel explicit and leaves the template-amplitude estimator unbiased, while affecting only the variance. This approximation does not change the conclusions of the forecast.

\section{The cold spot due to a cosmic texture}\label{sec:texture_per}

When CMB photons pass through the time-dependent gravitational potential of a collapsing texture, their temperature is either decreased (redshifted) or increased (blueshifted), depending on whether the photons pass before or after the collapse~\cite{sousa2013cmb}. To generate a CS, the texture must produce a temperature decrement through the non-linear Sachs--Wolfe, or Rees--Sciama, effect. This can be approximated as \citep{Farhang:2020gsa, Cruz:2007pe}:
\begin{equation}
   \label{texture}
\frac{\Delta T}{T}(\theta)=
\begin{cases}
-\epsilon \dfrac{1}{\sqrt{1+4 \dfrac{\theta^2}{\theta_{\mathrm{T}}^2}}} \quad\quad\quad\;\, \theta \leq \theta_* \;, \\[30pt]
-\dfrac{\epsilon}{2} e^{-\left(\theta^2-\theta_*^2\right) / 2 \theta_{\mathrm{T}}^2} \quad\quad \theta \geq \theta_* \;,
\end{cases} 
\end{equation}
where $\theta$ is the angle between the observed direction and the structure center, the amplitude is $\epsilon=8\pi^2 G\psi_0^2$, with $\psi_0$ being the energy scale of symmetry breaking and $\theta_* \equiv \sqrt{3}\,\theta_{\mathrm{T}}/ 2$. The definition of the scale parameter $\theta_T$ is
\begin{eqnarray}
    \theta_T=\frac{2\sqrt{2}k(1+z_{T})}{E(z_{T})\int_0^{z_{T}}\frac{d\bar{z}}{E(\bar{z})}},
\end{eqnarray}
where $k$ is a constant constrained by texture simulations, which suggest a value $\sim0.1$~\cite{Cruz:2007pe,Das:2008es}, $z_{T}$ is the redshift of the texture's center, and $E(z)=\left(\Omega_{\mathrm{m}}(1+z)^{3}+\Omega_{\Lambda}\right)^{1/2}$.

The texture not only shifts the temperature of the photons, but also changes their paths, thereby introducing a lensing signal. While void lensing acts as a diverging lens, the texture converges the photons according to the lens profile computed by \citep{DURRER1992527}. However, the approximation in \citep{DURRER1992527} is valid only for angles up to $\theta \approx\theta_T$. To obtain a profile applicable at larger angular distances, we follow \cite{vielva2010comprehensive, Farhang:2020gsa} and extend the profile from its half-maximum using a Gaussian function:
\begin{equation}
    \label{eq:texture_lensing}
\alpha(\theta)=
\begin{cases}
\dfrac{2\sqrt{2}\,\eps}{\thetac}
\,\dfrac{D_{LS}}{D_{S}}\,
\dfrac{\theta}{\sqrt{\,1+4\bigl(\theta/\thetac\bigr)^2}}\,, 
& \displaystyle \theta\le\thetas,\\[20pt]
\dfrac{\sqrt{2}\,\eps}{\thetac}
\,\dfrac{D_{LS}}{D_{S}}\,
\theta\,\exp\bigl[-(\theta^2-\thetas^2)/(2\thetac^{\,2})\bigr]\;,
& \displaystyle \theta>\thetas,
\end{cases}
\end{equation}
where $D_{\mathrm{S}}$ and $D_{\mathrm{LS}}$ are the comoving distances from the observer to the source and from the lens to the source, respectively. This extension ensures the continuity of both the profile and its first derivative.

\subsection{Temperature estimator}

\begin{figure}[t!]
    \centering
    \includegraphics[width=0.89\linewidth]{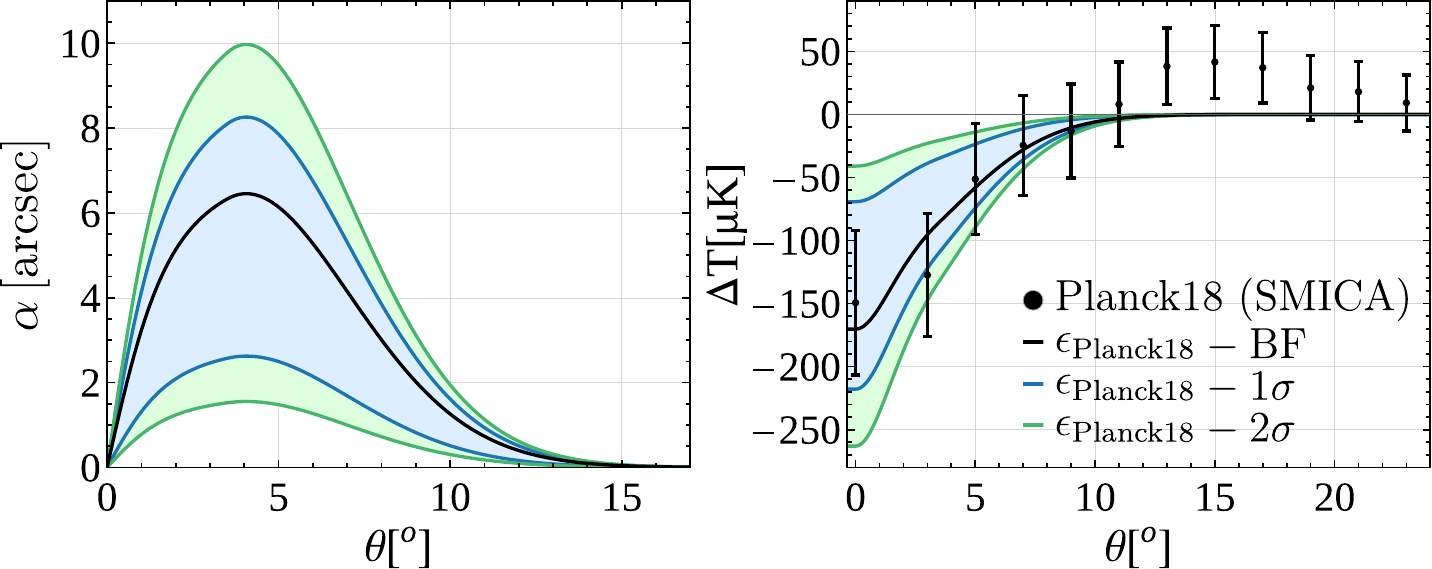}
    \caption{\textbf{Left:} Expected lensing angle profile ($\alpha$) in arcseconds for the estimated $\epsilon$ and $\theta_T$, best fit (BF), using \textit{Planck} 2018 (SMICA) data and the $\pm1\sigma$ and $\pm2\sigma$ regions. \textbf{Right}: The observed and expected temperature profile considering the same texture parameters.}\label{fig:texture_lensing_and_deltaT}
\end{figure}

For the temperature fit we use the following $\chi^2$ as a function of $\epsilon$ and $\theta_T$:
\begin{eqnarray}\label{chi_temperature}
    \chi^2(\epsilon, \theta_T)=\sum_i\bigg[\frac{\Delta T^{\rm TH}(\epsilon, \theta_T)-\Delta T^{\rm OBS}(\theta_i)}{\sigma_{\Delta T^{\rm OBS}(\theta_i)}}\bigg]^2.
\end{eqnarray} We compare the average temperature of a $2^\circ$ wide ring $\Delta T^{\rm OBS}(\theta_i)$, using the full-resolution SMICA map, centered at $\theta_i$ (as shown in Figure~\ref{fig:cs_position_and_profile}), with the expected texture temperature profile model $\Delta T^{\rm TH}(\epsilon, \theta_T)$. We also take into account the uncertainty from observational data $\sigma_{\Delta T^{\rm OBS}(\theta_i)}$. Using Eq.~\eqref{texture}, we obtain the temperature constraints from Planck 2018 data shown in Figure~\ref{fig:texture_lensing_and_deltaT}.

We estimate $\epsilon={6.25}^{+1.75}_{-3.71} \times10^{-5}$ and $\theta_T={4.1^{\circ}}^{+2.9^{\circ}}_{-1.7^{\circ}}$, using the priors $\epsilon \in [1,10]\times10^{-5}$ and $\theta_T\in [1^\circ,15^\circ]$, and considering $z_T\sim6$ as a good approximation for the center of the texture \cite{Cruz:2007pe,Farhang:2020gsa}. One could, of course, treat $z_T$ as a free parameter and vary $D_{LS}$ and $D_{S}$ in the estimator; however, for this first forecast we set $z_T$ fixed. From this point on we fix $\theta_T$ to its best-fit value obtained above, and we evaluate the S/N after marginalising over $\epsilon$ within its $\pm1\sigma$ and $\pm2\sigma$ ranges.

\section{Lensing effect in CMB polarization maps}

We generalize the temperature-based derivations of Refs.~\cite{Owusu:2022etl,Masina:2010dc} to include polarization. Since lensing remaps the fields as $\tilde X(\hat{\mathbf n})=X(\hat{\mathbf n}+\nabla\Theta)$, where $\Theta$ is the lensing potential, to first order we have
\begin{equation}\label{eq:deltaX_def}
\delta X(\hat{\mathbf n};\Theta)\equiv \tilde X(\hat{\mathbf n})-X(\hat{\mathbf n})
\simeq \nabla_i\Theta(\hat{\mathbf n})\,\nabla^i X(\hat{\mathbf n}),
\qquad X\in\{T,E,B\}.
\end{equation}
Equivalently, in harmonic space we write
$a^X_{\ell m}=a^{X(P)}_{\ell m}+a^{X(L_1)}_{\ell m}\equiv a^{X(P)}_{\ell m}+\delta a^X_{\ell m}(\Theta)$,
where $L_1$ denotes terms linear in $\Theta$. For convenience we define the spherical-harmonic coefficients of the lensing-potential template $b_{\ell m}$, which is considered here as axisymmetric such that all information is in $m=0$. The coefficient $b_{\ell0}$ can be written in terms of a normalized lensing profile $p(\theta)$ and the maximum value $\Theta_0$ of the lensing potential $\Theta(\theta)$\footnote{We consider a lensing profile with circular symmetry, as it only depends on $\theta$.}, as done in \cite{Owusu:2022etl}\footnote{$b_{\ell m}$ should not be confused with the instrumental beam window function often denoted by $b_\ell$.
} 
\begin{eqnarray}\label{bl0_1}
    b_{\ell 0} \equiv 2\pi\Theta_0\int d\theta \sin\theta   \, p(\theta) \, Y_{\ell 0}(\theta).
\end{eqnarray}
We can also define $\hat{b}_{\ell 0} \equiv  b_{\ell 0}/\Theta_0$, related to the lensing angle profile $\alpha(\theta)$ as
\begin{eqnarray}
    \Theta(\theta) =\int_0^{\theta}\alpha(\bar{\theta}) d\bar{\theta},
\end{eqnarray}


Lensing induces an off-diagonal covariance $\langle X_{\ell_1 m_1}Y_{\ell_2 m_2}\rangle$ proportional to the lensing potential multipoles. For any pair $X,Y\in\{T,E,B\}$ we define the off-diagonal estimator

\begin{equation}\label{FlmXY}
f^{\rm TH,XY}_{\ell_1\ell_2 m}
= 
\big\langle a^{X(P)*}_{\ell_1 m}\,a^{Y(L_1)}_{\ell_2 m}\big\rangle
+\big\langle a^{X(L_1)*}_{\ell_1 m}\,a^{Y(P)}_{\ell_2 m}\big\rangle.
\end{equation} 
For temperature, this has been demonstrated by \cite{Owusu:2022etl} to be
\begin{equation}\label{FTHTT}
\begin{split}
f_{\ell_1\ell_2m}^{TH(TT)}&=\big\langle a^{T}_{\ell_1m_1}a^{T(L_1)*}_{\ell_2m_2}\big\rangle + \big\langle a^{T*}_{\ell_1m_1}a^{T(L_1)}_{\ell_2 m_2}\big\rangle \\&= \Theta_0 (-1)^m \sum_{\ell_3} {}^0G_{\ell_1 \ell_2 \ell_3}^{-m\,m\,0}
\bigg(
C_{\ell_1}^{(P)} \beta_1 + C_{\ell_2}^{(P)} \beta_2
\bigg)\hat{b}_{\ell_3 0}\,.
\end{split}
\end{equation}
where the related spin-0 Gaunt integral is given as
\begin{equation}
\begin{split}
{}^0G_{\ell_1\ell_2\ell_3}^{-mm0}=\sqrt{\frac{(2\ell_1+1)(2\ell_2+1)(2\ell_3+1)}{4\pi}}
\begin{pmatrix}
\ell_1 & \ell_2 & \ell_3\\    
0 & 0 & 0
\end{pmatrix}
\begin{pmatrix}
\ell_1 & \ell_2 & \ell_3\\
-m & m & 0
\end{pmatrix}
\end{split},
\end{equation}
and coefficients $\beta_1$ and $\beta_2$ are

\begin{equation}
    \beta_1 = \frac{\ell_1(\ell_1+1)-\ell_2(\ell_2+1)+\ell_3(\ell_3+1)}{2} \;\;\;\mathrm{and}\;\;\; \beta_2 = \frac{\ell_2(\ell_2+1)-\ell_1(\ell_1+1)+\ell_3(\ell_3+1)}{2}
\end{equation}

We now write the corresponding first-order expressions for $EE$, $ET$ ($TE$), $EB$ ($BE$), and $BT$ ($TB$)\footnote{Even if $C_\ell^{\rm BT}=0$, lensing generates an off-diagonal $\mathrm{TB}$ signal through
\begin{equation}
\left\langle \tilde{T}\,\tilde{B}\right\rangle_{\ell_1\neq\ell_2}
\simeq
\left\langle T\,\delta B\right\rangle
+
\left\langle \delta T\,B\right\rangle,
\end{equation}
where tildes denote lensed quantities. $\langle T\,\delta B\rangle$ is proportional to $C_\ell^{TE}$, while $\langle \delta T\,B\rangle$ is proportional to $C_\ell^{TB}$ (vanishing in parity-even cosmology).} (the $BB$ response vanishes in the negligible primordial-$B$ limit). The first-order off-diagonal estimators, i.e. the cross-correlations between all pairs of observables, are
\vspace{-0.1cm}
\begin{eqnarray}\label{FTHEE}
f_{\ell_1\ell_2m}^{\rm TH(EE)} =&\hspace{-0.93cm}\hspace{-2.8cm}\Theta_0\sum_{\ell_3}{}^2G_{\ell_1\ell_2\ell_3}^{-mm0}\mu_+\bigg(C_{\ell_1}^{EE}\beta_1+ C_{\ell_2}^{EE}\beta_2\bigg)\hat{b}_{\ell_3 0},  \\
f^{\rm TH(ET)}_{\ell_1 \ell_2 m} \label{FTHET}=&\hspace{-0.01cm}\Theta_0 \Bigg[C^{\rm ET}_{\ell_2}\sum_{\ell_3}{}^2G_{\ell_1\ell_2\ell_3}^{-m m 0} \mu_+ \beta_2\hat{b}_{\ell_3 0} +(-1)^m C^{\rm ET}_{\ell_1}\sum_{\ell_3}{}^0G_{\ell_1\ell_2\ell_3}^{-m m 0} \beta_1 \hat{b}_{\ell_3 0}\Bigg] \,, \\
f^{\rm TH(EB)}_{\ell_1\ell_2 m}\label{FTH_EB_final}
=&\hspace{-5.65cm}\,\Theta_0 
\sum_{\ell_3} {}^{2}G_{\ell_1\ell_2\ell_3}^{-m\,m\,0}\,\mu_-\,
C_{\ell_1}^{\rm EE}\,\beta_1 \,\hat{b}_{\ell_3 0}, \\
f^{\rm TH(BT)}_{\ell_1\ell_2 m} \label{FTH_BT_final}
=& \hspace{-5.55cm}\Theta_0\,C_{\ell_2}^{\rm ET}
\sum_{\ell_3} {}^{2}G_{\ell_1\ell_2\ell_3}^{-m\,m\,0}\,\mu_-\,
\beta_2\,\hat{b}_{\ell_3 0}\,.
\end{eqnarray}

with $\mu_+ = (1+(-1)^L)(-1)^m/2$, $\mu_- = (1-(-1)^L)(-1)^m/2$, $L=\ell_1+\ell_2+\ell_3$ and where the spin-2 Gaunt integral is given as
\begin{equation}
\begin{split}
{}^2G_{\ell_1\ell_2\ell_3}^{-mm0}=\sqrt{\frac{(2\ell_1+1)(2\ell_2+1)(2\ell_3+1)}{4\pi}}
\begin{pmatrix}
\ell_1 & \ell_2 & \ell_3\\    
2 & -2 & 0
\end{pmatrix}
\begin{pmatrix}
\ell_1 & \ell_2 & \ell_3\\
-m & m & 0
\end{pmatrix} \, ,
\end{split}
\end{equation}

Here we have used the fact that, for a primordial isotropic and Gaussian signal, the two-point correlation functions take the form: $\big\langle a_{\ell_1 m_1}^{\rm X} a_{\ell_2 m_2}^{\rm X*}\big\rangle = \delta_{\ell_1 \ell_2}\delta_{m_1 m_2} C_{\ell_1}^{\rm XX}$. The Kronecker deltas make it diagonal in the $m$ index. The diagonal sector $\ell_1=\ell_2$ requires some care: individual $m$ modes can respond to an axisymmetric lensing template, but the first-order contribution to the isotropically averaged diagonal power cancels after the full $m$ sum. We therefore build the estimator from the clean off-diagonal covariance, taking $\ell_2=\ell_1+1,\ldots,\ell_1+\Delta\ell$. Although each individual coupling is weak, their broad range still yields a statistically significant contribution. It was shown in~\cite{Owusu:2022etl} that most of the signal is contained in $|\ell_1-\ell_2|\equiv \Delta\ell\leq 40$.

The appendices give the complete derivations for the $EE$, $TE$ ($ET$), $TB$ ($BT$), $EB$ ($BE$) and $BB$ channels. See \cite{Masina:2009wt,Masina:2010dc,Owusu:2022etl} for the $TT$ cross-correlations and variance derivation\footnote{A full GMV implementation in our template-amplitude framework would correspond to replacing the block-diagonal (per-$XY$) weighting by the inverse of the full joint covariance in the space of quadratic pairs ($TT$, $EE$, $ET$, $EB$ and $TB$) for each $(\ell_1,\ell_2,m)$ configuration, as discussed in~\cite{Maniyar:2021msb}.}. The physical idea is that, in the absence of a localized lensing profile, statistical isotropy makes the CMB covariance diagonal in harmonic space. A texture-like lensing profile produces a deterministic pattern of off-diagonal mode couplings, and each configuration $(XY,\ell_1,\ell_2,m)$ is therefore a noisy measurement of the same template amplitude.

For an ordered pair $XY$, the Gaussian variance of this off-diagonal quantity is
\begin{equation}
{\rm Var}\!\left[f^{XY}_{\ell_1\ell_2m}\right]
=\frac{1}{2}(1+\delta_{m0})\,
\mathfrak C_{\ell_1}^{XX}\mathfrak C_{\ell_2}^{YY},
\end{equation}
where $X$ labels the field at $\ell_1$ and $Y$ the field at $\ell_2$. The factor $(1+\delta_{m0})/2$ follows from working with real combinations of the $m$ and $-m$ modes: modes with $m>0$ have a corresponding $-m$ partner, while $m=0$ is counted only once. Equivalently, the inverse-variance sum carries the factor $(2-\delta_{m0})$. We then obtain the corresponding S/N as
\begin{equation}\label{StoN}
\left(\frac{S}{N}\right)^2_{XY}
=\sum_{\ell_1=2}^{\ell_{\rm max}}
\sum_{\ell_2=\ell_1+1}^{\ell_1+\Delta\ell}
\sum_{m=0}^{\ell_1}
(2-\delta_{m0})
\frac{
\big[f_{\ell_1 \ell_2 m}^{\rm{TH},XY}\big]^2
}{
\mathfrak{C}_{\ell_1}^{XX}\mathfrak{C}_{\ell_2}^{YY}
}\,.
\end{equation}
Here
\begin{equation}
\mathfrak{C}_\ell^{XX}\equiv
f_{\rm sky}^{-1/2}
\left(\tilde{C}_\ell^{XX}+N_\ell^{XX}\right)
\end{equation}
is the effective diagonal covariance spectrum of the observed field $X$,
including the lensed CMB spectrum, instrumental noise and beam
deconvolution through $N_\ell^{XX}$.

\subsection{Lensing estimator} \label{sec:lens_estimator}

By comparing the expected and observed off-diagonal $TT$ lensing correlations, we obtained a $\chi^2$ estimator in \cite{Owusu:2022etl}. Here we extend the same matched-filter logic to a general ordered $XY$ pair, adapted to the texture case, to measure the texture parameters $\epsilon$ and $\theta_T$:
\begin{eqnarray}\label{chisq_lensing}   
   \chi_{XY}^2(\epsilon,\theta_T) =
   \sum_{\ell_1 = 2}^{\ell_{\rm max}}
   \sum_{\ell_2=\ell_1 + 1}^{\ell_1 + \Delta \ell}
   \sum_{m=0}^{\ell_1}
   (2-\delta_{m0})
   \frac{
   \left[
   f^{\rm OBS,XY}_{\ell_1 \ell_2 m}
   -
   f^{\rm TH,XY}_{\ell_1 \ell_2 m}(\epsilon,\,\theta_T)
   \right]^2
   }{
   \mathfrak{C}_{\ell_1}^{XX}
   \mathfrak{C}_{\ell_2}^{YY}
   } \; ,
\end{eqnarray}
where $f^{\rm OBS,XY}_{\ell_1 \ell_2 m}$ is the observed off-diagonal two-point correlation defined by Eqs.~\eqref{FTHTT}--\eqref{FTH_BT_final}. For fixed profile shape, $b_{\ell0}=\Theta_0\hat b_{\ell0}$, and therefore $f^{\rm TH,XY}_{\ell_1\ell_2m}$ is linear in the template amplitude. Equation~\eqref{chisq_lensing} is thus the inverse-variance matched filter for that amplitude. In the forecast, the Fisher information is the sum of the squared template response divided by its Gaussian variance, which gives Eq.~\eqref{StoN}; consequently the S/N scales linearly with $\Theta_0$, or equivalently with $\epsilon$ once $\theta_T$ is fixed.
\section{Forecast for the Simons Observatory}

To forecast the sensitivity of the Simons Observatory to temperature and polarization lensing, we compute Eq.~\eqref{StoN} with the texture lensing profile. For this estimate, we use the specifications provided for the updated optimal case of the Simons Observatory Large Aperture Telescope Survey \cite{SimonsObservatory:2025wwn}, listed in Table~\ref{tab:specs}. This corresponds to a co-added temperature noise level of $2.6\,\mu\mathrm{K}\cdot \mathrm{arcmin}$ after combining all frequencies, over an observed sky fraction $f_{\rm sky}=0.61$. 

\begin{table}[ht]
    \centering
    \begin{tabular}{ccc}
        \hline
        Frequency [GHz] & FWHM [arcmin] & Goal Depth [$\mu\mathrm{K}\cdot \mathrm{arcmin}$] \\
        \hline
        27  & 7.4 & 44  \\
        39  & 5.1 & 23  \\
        93  & 2.2 & 3.8 \\
        145 & 1.4 & 4.1 \\
        225 & 1.0 & 10  \\
        280 & 0.9 & 25  \\
        \hline
    \end{tabular}\caption{Anticipated instrumental specifications and map-depth characteristics for the full nine-year SO LAT survey (2025--2034) following Ref.~\cite{SimonsObservatory:2025wwn}.}\label{tab:specs}
\end{table}
We model the noise power spectrum as
\begin{equation}
N_{\ell} \;=\; \left[\sum_{i=1}^{\mathrm{\#\,freqs}}\left(\sigma_{\theta,i}^{2}\,\sigma_{T,i}^{2}\,
\exp\!\left[\frac{\ell(\ell+1)\,\theta_{i}^{2}}{8\log 2}\right]\right)^{-2}\right]^{-1/2}.
\end{equation}
where $\sigma_{\theta,i}$ is the FWHM of frequency channel $i$ and $\sigma_{T,i}$ is its depth, or sensitivity. For the polarization sensitivity we set $\sigma_{P,i} \equiv \sqrt{2} \sigma_{T,i}$, assuming uncorrelated noise between the $Q$ and $U$ maps. The final effective spectrum $\mathfrak{C}_{\ell}$ used in the covariance is
\begin{equation}
\mathfrak{C}_{\ell} \;\equiv\; \frac{1}{\sqrt{f_{\mathrm{sky}}}}\bigl(\tilde{C}_{\ell}+N_{\ell}\bigr).
\end{equation}
In the numerical forecast we distinguish two response prescriptions. The ``lensed response'' case, used as our main result, uses lensed CMB spectra both in the covariance above and in the response functions. This choice better reflects the fact that the observed CMB maps are already lensed by standard large-scale structure, and provides a conservative estimate of the template S/N. We also show a ``first-order response'' comparison, where the response functions use unlensed spectra while the covariance still uses lensed spectra; this corresponds most directly to the perturbative derivation in which the localized texture lenses the primary CMB. The difference between the two curves quantifies the sensitivity of the forecast to the treatment of standard lensing in the response. We take $\ell_{\rm max}=4250$ for all channels shown in the final forecast and limit $\Delta\ell=37$, since larger separations add negligible signal for the adopted texture profile and survey specifications.

As long as the mask does not cover a significant part of the lensing profile $p(\theta)$ (which is encoded in harmonic space by $b_{\ell0}$), it will not substantially affect the S/N. Thus, a mask located more than $\sim20^\circ$ from the Cold Spot center should not significantly reduce the estimator sensitivity. A mask can still affect the signal slightly by introducing mode coupling that attenuates the expected lensing footprint, but realistic simulations from our previous work indicate that this is a subdominant effect \cite{Owusu:2022etl}.

\begin{figure}[h!]
    \centering
    \includegraphics[width=1\linewidth]{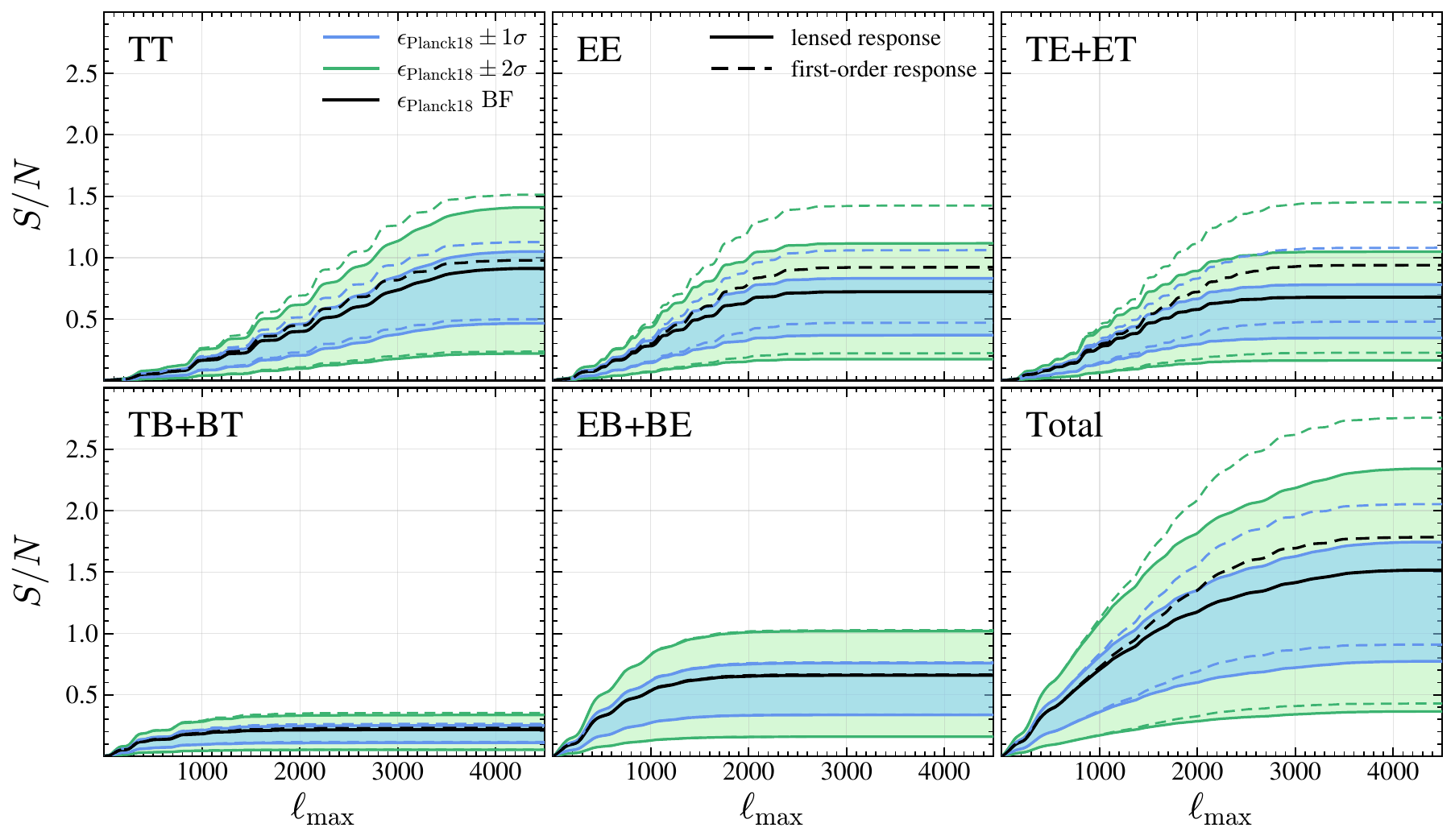}
    \caption{Expected signal-to-noise ratio (S/N) for the lensing signal in Simons Observatory data using our estimator on $TT$, $EE$, $TE+ET$, $TB+BT$, $EB+BE$ and their total combination, obtained by adding the ordered-pair contributions in quadrature. We consider the best fit (BF) $\epsilon$ and $\theta_T$, and the $\pm1\sigma$ and $\pm2\sigma$ bounds for the $\epsilon$ texture parameter obtained using \textit{Planck} 2018 (SMICA) data \cite{Akrami:2018mcd}. Solid curves show the lensed-response forecast, while dashed curves show the first-order response.}\label{fig:texture_s2n}
\end{figure}

\begin{figure}[h!]
    \centering
    \includegraphics[width=0.82\linewidth]{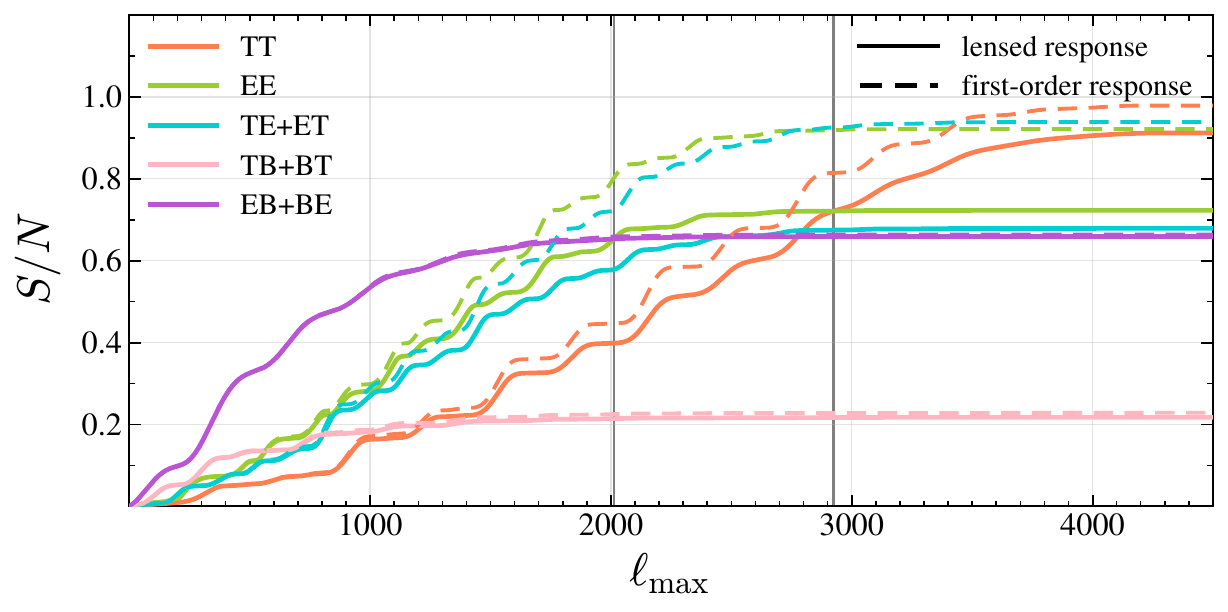}
    \caption{Best-fit cumulative signal-to-noise ratio for the individual quadratic pairs. Solid curves show the lensed-response forecast, while dashed curves show the first-order response. The vertical black lines mark the cumulative scales at which the strongest best-fit lensed-response channel changes: $EB+BE$ dominates before the first vertical line, $EE$ surpasses it over the intermediate range, and $TT$ dominates at higher $\ell_{\rm max}$.}\label{fig:texture_s2n_bf}
\end{figure}

Using this estimator and the survey specifications above, the estimated S/N for $TT$, $EE$, $TE+ET$, $TB+BT$, $EB+BE$ and the total lensing signal using \textit{Planck} constraints for $\epsilon$ and $\theta_T$ is given in Figure~\ref{fig:texture_s2n}. Figure~\ref{fig:texture_s2n_bf} highlights that the relative importance of the polarization estimators is scale dependent: the $EB+BE$ channel dominates at low cumulative multipoles, as expected by analogy with standard CMB lensing, but the cumulative hierarchy changes once the full multipole range and the \(BB\) covariance are included, for the adopted SO noise model.

For the optimistic scenario in which the texture amplitude is set by the $2\sigma$ upper limit on $\epsilon$ from \textit{Planck} 2018 data, the combined temperature-plus-polarization estimator gives a $2.3\sigma$ detection of the texture lensing signature. This corresponds to detecting a lensing profile with an average deflection angle below $\sim10^{\prime\prime}$. In the same scenario, the $TT$ channel alone gives a $1.4\sigma$ detection, while $EE$, $TE+ET$, $TB+BT$ and $EB+BE$ contribute $1.1\sigma$, $1.1\sigma$, $0.3\sigma$ and $1.0\sigma$, respectively. For the best-fit texture parameters, the expected total S/N is $1.5$, still notable given the expected typical deflection angle of $\lesssim6^{\prime\prime}$. For the first-order response comparison, in which the response functions use unlensed spectra while the covariance still uses lensed spectra, the total S/N is $1.8$ for the best-fit texture parameters and $2.8$ for the optimistic $2\sigma$ upper-limit case. Thus, using the lensed response lowers the cumulative S/N by about $17$--$18\%$ relative to the first-order response. This quantifies the impact of standard LSS lensing on the response spectra. These forecasts are limited
mainly by the precision with which $\epsilon$ and $\theta_T$ are currently
inferred from \textit{Planck} 2018 temperature data
\cite{Akrami:2018mcd}.

\section{Conclusions and Perspectives}

Cosmic textures constitute a unique messenger from the very early universe, tracing physics at energy scales of $10^{15}\!-\!10^{16}\,\mathrm{GeV}$ characteristic of Grand Unified Theories (GUTs)~\cite{Turok:1990gw}. Their imprints on the CMB therefore encode information from an era earlier than photon decoupling and, potentially, earlier than inflation itself, providing an observational handle on GUT-scale physics that is otherwise inaccessible to laboratory experiments or lower-energy cosmological probes \cite{Turok:1990gw,DURRER1992527,Durrer:2001cg}. Detecting (or tightly constraining) a texture signature would thus illuminate the structure of the primordial universe and place direct limits on high-energy particle models far beyond the reach of colliders.

The lensing signal from a texture capable of producing the CS could be detected at $2.3\sigma$ significance with the Simons Observatory, in the optimistic scenario in which the texture amplitude equals the $2\sigma$ upper-bound constraint derived from \textit{Planck} 2018 data \cite{Akrami:2018mcd}. For the best-fit texture parameters, the S/N is reduced to $1.5$, still a notable sensitivity given the expected bending of no more than $6^{\prime\prime}$. These numbers highlight the central role of polarization. Although \textit{Planck} polarization is noise-dominated beyond $\ell\!>\!1000$ \cite{Planck:2019nip}, next-generation surveys will reach $\ell\!>\!3000$ in polarization. In our SO forecast, adding polarization improves the cumulative S/N by $\sim67\%$ relative to temperature alone, making weak localized lensing signals accessible that would otherwise remain below the temperature-only sensitivity.

The main methodological contribution of this work is a targeted template-amplitude pipeline: instead of reconstructing an arbitrary lensing field, it projects the standard quadratic lensing response onto a localized profile and combines the ordered temperature and polarization pairs. Although a topological defect served as our case study, the pipeline is agnostic to the physical origin of the lensing profile and can be applied to any localized imprint with a specified morphology and center.

Looking several decades ahead, a next-generation facility, such as the proposed CMB-HD \cite{CMB-HD:2022bsz}, reaching $\ell\sim25000$, could detect even fainter phenomena. With that angular resolution, the experiment will push the sensitivity frontier to phenomena an order of magnitude fainter than those accessible today, enabling direct detection of the Kaiser--Stebbins effect and lensing imprint from cosmic strings, whose characteristic lensing deflection scale is $\lesssim1''$ \cite{Planck:2013mgr}.

By combining temperature and polarization lensing imprints, the proposed estimator provides a promising route to put not only the Cold Spot under the lens, but also other faint localized lensing footprints in future CMB data.   

\appendix\label{sec:app}
\section{The $f^{EE}_{\ell_1\ell_2m}$ component}\label{FEE_complete}

Following the temperature derivation of Refs.~\cite{Owusu:2022etl, Masina:2010dc, Masina:2009wt}, we expand the polarization field in spin-weighted spherical harmonics~\cite{Hu:2000ee}:
\begin{eqnarray}
{}_\pm X(\hat{n})=\sum_{\ell m}{}_\pm X_{\ell m} {}_{\pm 2}Y^m_{\ell}(\hat{n})\,,
\end{eqnarray}
where ${}_\pm X$ denotes the complex Stokes combination
\begin{eqnarray}
{}_\pm X=Q(\hat{n})\pm iU(\hat{n}) \, ,
\end{eqnarray}
which is a spin-2 field. With the parity transformation ${}_SY^m_\ell \rightarrow (-1)^\ell {}_{-S}Y^m_\ell$, we define the parity eigenstates~\cite{kamionkowski1997statistics}
\begin{eqnarray}\label{X_E_B}
{}_\pm X_{\ell m}=a^{\rm E}_{\ell m} \pm ia^{\rm B}_{\ell m}\,,
\end{eqnarray}
so that $E(\hat{n})$ and $B(\hat{n})$ are real scalar quantities with even ($(-1)^\ell$, electric) and odd ($(-1)^{\ell+1}$, magnetic) parity, respectively:
\begin{eqnarray}
E(\hat{n})=\sum_{\ell m}a^{\rm E}_{\ell m}Y^m_{\ell}(\hat{n}),\;\;\;\;\;\; a^{\rm E}_{\ell m}=-\frac{a_{2\ell m}+a_{-2\ell m}}{2} ,
\end{eqnarray}
\begin{eqnarray}
B(\hat{n})=\sum_{\ell m}a^{\rm B}_{\ell m}Y^m_{\ell}(\hat{n}),\;\;\;\;\;\; a^{\rm B}_{\ell m}=i\frac{a_{2\ell m}-a_{-2\ell m}}{2}.
\end{eqnarray}

Weak lensing remaps the primordial anisotropy and induces off-diagonal correlations in harmonic space. To first order in the lensing-potential multipoles $b_{\ell m}$, the spin fields transform as~\cite{Hu:2000ee}
\begin{eqnarray}\label{X_harmonics}
{}_\pm\Tilde{X}_{\ell m}={}_\pm X_{\ell m}+ \sum_{\ell_1 m_1}\sum_{\ell_2 m_2}b_{\ell_1 m_1}{}_\pm X_{\ell_2 m_2} {}_{\pm 2}I^{m m_1 m_2}_{\ell \ell_1 \ell_2},
\end{eqnarray}
where we keep only terms linear in $b_{\ell m}$. The geometrical coupling is
\begin{eqnarray}
{}_{\pm 2}I^{m m_1 m_2}_{\ell \ell_1 \ell_2} = \int d\hat{n} {}_{\pm 2}Y^{m*}_\ell(\nabla_i Y^{m_1}_{\ell_1})(\nabla^i {}_{\pm 2}Y^{m_2}_{\ell_2}),
\end{eqnarray}
Substituting Eq.~\eqref{X_E_B} into Eq.~\eqref{X_harmonics} gives
\begin{eqnarray}
{}_\pm\Tilde{X}_{\ell m}={}_\pm X_{\ell m}+ \sum_{\ell_1 m_1}\sum_{\ell_2 m_2}b_{\ell_1 m_1}(a^{\rm E}_{\ell_2 m_2} \pm ia^{\rm B}_{\ell_2 m_2}) {}_{\pm 2}I^{m m_1 m_2}_{\ell \ell_1 \ell_2} \,.
\end{eqnarray}
Using the spin-2 angular Laplacian
\begin{eqnarray}
\nabla^2{}_{\pm 2}Y^m_\ell=[-\ell(\ell+2)+4]{}_{\pm 2}Y^m_\ell,
\end{eqnarray}
one obtains
 \begin{eqnarray}
\!\!\!\!{}_{\pm 2}I^{m m_1 m_2}_{\ell \ell_1 \ell_2} =\frac{1}{2}[\ell_1(\ell_1+1)+\ell_2(\ell_2+1)-\ell(\ell+1)]\int d\hat{n} {}_{\pm 2}Y^{m*}_\ell(\nabla_i Y^{m_1}_{\ell_1})(\nabla^i {}_{\pm 2}Y^{m_2}_{\ell_2}) \,,
\end{eqnarray}
which leads to
 \begin{equation}
\begin{split}
 {}_{\pm 2}I^{m m_1 m_2}_{\ell \ell_1 \ell_2}&=\frac{1}{2}[\ell_1(\ell_1+1)+\ell_2(\ell_2+1)-\ell(\ell+1)] \sqrt{\frac{(2\ell_1+1)(2\ell_2+1)(2\ell+1)}{4 \pi}}\times\\&
\begin{pmatrix}
\ell_1 & \ell_2 & \ell_3\\    
2 & 0 & -2
\end{pmatrix}
\begin{pmatrix}
\ell_1 & \ell_2 & \ell_3\\
-m_1 & m_2 & m
\end{pmatrix}
(-1)^{m_1+2}.
\end{split}
\end{equation}
The scalar $E$ and pseudo-scalar $B$ harmonic coefficients can equivalently be written as
\begin{equation}
\begin{split}
& a^{\rm E}_{\ell m}=\frac{1}{2}({}_2X_{\ell m}+{}_{-2}X_{\ell m}) \;\;\;\; \textrm{and} \;\;\;\; a^{\rm B}_{\ell m}=\frac{i}{2}({}_{-2}X_{\ell m}-{}_{2}X_{\ell m}),
\end{split}
\end{equation}
These relations allow us to express both the $EE$ component and the crossed $E$--$T$ spectra in terms of spin fields.
\begin{eqnarray} \label{EE_1st}
\big\langle a^{\rm E}_{\ell m}a^{\rm E *}_{\ell' m'}\big\rangle=\frac{1}{4}[{}_2\tilde{X}_{\ell m}{}_2\tilde{X}_{\ell' m'}^*+ {}_{-2}\tilde{X}_{\ell m}{}_{-2}\tilde{X}_{\ell' m'}^*+{}_{-2}\tilde{X}_{\ell m}{}_{2}\tilde{X}_{\ell' m'}^*+{}_{2}\tilde{X}_{\ell m}{}_{-2}\tilde{X}_{\ell' m'}^*].
\end{eqnarray}
Keeping only terms linear in $b_{\ell m}$, the four contributions in Eq.~\eqref{EE_1st} are
\begin{equation}
\begin{split}
{}_2\tilde{X}_{\ell m}{}_2\tilde{X}_{\ell' m'}^*&=\sum_{\ell_1 m_1}\sum_{\ell_2 m_2}b_{\ell_1 m_1}\big\langle{}_2X_{\ell_2 m_2}{}_2X_{\ell' m'}^*\big\rangle {}_{2}I^{m m_1 m_2}_{\ell \ell_1 \ell_2}
\\&=\sum_{\ell_1 m_1}\sum_{\ell_2 m_2}b_{\ell_1 m_1}\big\langle(a^{\rm E}_{\ell_2 m_2} + ia^{\rm B}_{\ell_2 m_2})(a^{\rm E}_{\ell' m'} - ia^{\rm B}_{\ell' m'})\big\rangle {}_{2}I^{m m_1 m_2}_{\ell \ell_1 \ell_2} \\& =\sum_{\ell_1 m_1}\sum_{\ell_2 m_2}b_{\ell_1 m_1}(C_{\ell_2}^{EE}+C_{\ell_2}^{BB}) {}_{2}I^{m m_1 m_2}_{\ell \ell_1 \ell_2}\delta_{\ell_2\ell'}\delta_{m_2m'} \, ,
\end{split}
\end{equation}

\begin{equation}
\begin{split}
{}_{-2}\tilde{X}_{\ell m}{}_{-2}\tilde{X}_{\ell' m'}^*&=\sum_{\ell_1 m_1}\sum_{\ell_2 m_2}b_{\ell_1 m_1}\big\langle{}_{-2}X_{\ell_2 m_2}{}_{-2}X_{\ell' m'}^*\big\rangle {}_{-2}I^{m m_1 m_2}_{\ell \ell_1 \ell_2}
\\&=\sum_{\ell_1 m_1}\sum_{\ell_2 m_2}b_{\ell_1 m_1}(a^{\rm E}_{\ell_2 m_2} - ia^{\rm B}_{\ell_2 m_2})(a^{\rm E}_{\ell' m'} + ia^{\rm B}_{\ell' m'}) {}_{-2}I^{m m_1 m_2}_{\ell \ell_1 \ell_2} \\& =\sum_{\ell_1 m_1}\sum_{\ell_2 m_2}b_{\ell_1 m_1}(C_{\ell_2}^{EE}+C_{\ell_2}^{BB}) {}_{-2}I^{m m_1 m_2}_{\ell \ell_1 \ell_2}\delta_{\ell_2\ell'}\delta_{m_2m'} \, ,
\end{split}
\end{equation}

\begin{equation}
\begin{split}
{}_{-2}\tilde{X}_{\ell m}{}_{2}\tilde{X}_{\ell' m'}^*&=\sum_{\ell_1 m_1}\sum_{\ell_2 m_2}b_{\ell_1 m_1}\big\langle{}_{-2}X_{\ell_2 m_2}{}_{2}X_{\ell' m'}^*\big\rangle {}_{-2}I^{m m_1 m_2}_{\ell \ell_1 \ell_2}
\\&=\sum_{\ell_1 m_1}\sum_{\ell_2 m_2}b_{\ell_1 m_1}\big\langle(a^{\rm E}_{\ell_2 m_2} - ia^{\rm B}_{\ell_2 m_2})(a^{\rm E}_{\ell' m'} - ia^{\rm B}_{\ell' m'})\big\rangle {}_{-2}I^{m m_1 m_2}_{\ell \ell_1 \ell_2} \\& =\sum_{\ell_1 m_1}\sum_{\ell_2 m_2}b_{\ell_1 m_1}(C_{\ell_2}^{EE}-C_{\ell_2}^{BB}) {}_{-2}I^{m m_1 m_2}_{\ell \ell_1 \ell_2}\delta_{\ell_2\ell'}\delta_{m_2m'} \, ,
\end{split}
\end{equation}
\begin{equation}
\begin{split}
{}_{2}\tilde{X}_{\ell m}{}_{-2}\tilde{X}_{\ell' m'}^*&=\sum_{\ell_1 m_1}\sum_{\ell_2 m_2}b_{\ell_1 m_1}\big\langle{}_{2}X_{\ell_2 m_2}{}_{-2}X_{\ell' m'}^*\big\rangle {}_{2}I^{m m_1 m_2}_{\ell \ell_1 \ell_2}
\\&=\sum_{\ell_1 m_1}\sum_{\ell_2 m_2}b_{\ell_1 m_1}\big\langle(a^{\rm E}_{\ell_2 m_2} - ia^{\rm B}_{\ell_2 m_2})(a^{\rm E}_{\ell' m'} - ia^{\rm B}_{\ell' m'})\big\rangle {}_{2}I^{m m_1 m_2}_{\ell \ell_1 \ell_2} \\& =\sum_{\ell_1 m_1}\sum_{\ell_2 m_2}b_{\ell_1 m_1}(C_{\ell_2}^{EE}-C_{\ell_2}^{BB}) {}_{2}I^{m m_1 m_2}_{\ell \ell_1 \ell_2}\delta_{\ell_2\ell'}\delta_{m_2m'}\;\;. 
\end{split}
\end{equation}
Combining these terms and aligning the $\hat z$ axis with the center of the lensing profile, so that only $m_1=0$ contributes, gives
\begin{eqnarray}
\big\langle a^{E}_{\ell m}a^{E(L_1)*}_{\ell'm'}\big\rangle=\frac{1}{2}C_{\ell'}^{EE}\sum_{\ell_1}b_{\ell_1 0} \hspace{1mm} {}_{2}I^{m 0 m'}_{\ell \ell_1 \ell'}(1+(-1)^L) \, ,
\end{eqnarray}
\begin{eqnarray}
\big\langle a^{E(L_1)}_{\ell'm'}a^{E*}_{\ell m}\big\rangle=\frac{1}{2}C_{\ell}^{EE}\sum_{\ell_1}b_{\ell_1 0} \hspace{1mm} {}_{2}I^{m' 0 m}_{\ell' \ell_1 \ell}(1+(-1)^L) \, ,
\end{eqnarray}
where we have used
\begin{eqnarray}\label{Ipm2}
{}_{\pm 2}I^{m m_1 m_2}_{\ell \ell_1 \ell_2} =(-1)^L {}_{\mp 2}I^{m m_1 m_2}_{\ell \ell_1 \ell_2} ,
\end{eqnarray}
with $L=\ell+\ell_1+\ell_2$.
The resulting theoretical off-diagonal two-point function is therefore
\begin{equation}
\begin{split}
f_{\ell_1\ell_2m}^{TH(EE)}&=\big\langle a^{E}_{\ell_1m_1}a^{E(L_1)*}_{\ell_2m_2}\big\rangle + \big\langle a^{E(L_1)}_{\ell_1m_1}a^{E*}_{\ell_2 m_2}\big\rangle \\& =\frac{(-1)^{m}}{2}\sum_{\ell_3}{}^2G_{\ell_1\ell_2\ell_3}^{-mm0}(1+(-1)^L)\bigg(C_{\ell_1}^{EE}\frac{\ell_1(\ell_1+1)-\ell_2(\ell_2+1)+\ell_3(\ell_3+1)}{2}+ \\&C_{\ell_2}^{EE}\frac{\ell_2(\ell_2+1)-\ell_1(\ell_1+1)+\ell_3(\ell_3+1)}{2}\bigg)b_{\ell_3 0} \, ,
\end{split}
\end{equation}
where the spin-2 Gaunt integral is
\begin{equation}
\begin{split}
{}^2G_{\ell_1\ell_2\ell_3}^{-mm0}=\sqrt{\frac{(2\ell_1+1)(2\ell_2+1)(2\ell_3+1)}{4\pi}}
\begin{pmatrix}
\ell_1 & \ell_2 & \ell_3\\    
2 & 0 & -2
\end{pmatrix}
\begin{pmatrix}
\ell_1 & \ell_2 & \ell_3\\
-m & 0 & m
\end{pmatrix}
\end{split}.
\end{equation}

\section{Variance of $f^{EE}_{\ell_1\ell_2m}$}\label{VarEE}

We define the variance as $\sigma^2_{\ell_1\ell_2m}=\langle f^2_{\ell_1\ell_2m}\rangle-\langle f_{\ell_1\ell_2m}\rangle^2$. For the $EE$ case it is useful to introduce
\begin{eqnarray}
    F^{EE}_{\ell_1\ell_2m} \equiv a^{E*}_{\ell_1m}a^{E}_{\ell_2m},
\end{eqnarray}
\begin{eqnarray}
    F^{EE}_{\ell_1\ell_2-m}=a^{E*}_{\ell_1-m}a^{E}_{\ell_2-m} \, ,
\end{eqnarray}
where 
\begin{eqnarray}
a^{\rm E}_{\ell m} = a_{\ell m}^{E(P)}+a_{\ell m}^{E(L)} .
\end{eqnarray}
For the real and imaginary combinations we use
\begin{eqnarray}
f_{\ell_1 \ell_2 m}\equiv \frac{1}{2}\left(F_{\ell_1 \ell_2 m}+F_{\ell_1 \ell_2 -m}\right), \qquad
g_{\ell_1 \ell_2 m}\equiv \frac{1}{2i}\left(F_{\ell_1 \ell_2 m}-F_{\ell_1 \ell_2 -m}\right).
\end{eqnarray}
	\begin{eqnarray}
	\!\!\!\!\!\!\big\langle f_{\ell_1 \ell_2 m} f_{\ell_1' \ell_2' m'}\big\rangle\!=\!\frac{1}{4} \big\langle \!F_{\ell_1 \ell_2 m}F_{\ell'_1 \ell'_2 m'}\!+\!F_{\ell_1 \ell_2 m}F_{\ell'_1 \ell'_2 -m'}\!+\!F_{\ell_1 \ell_2 -m}F_{\ell'_1 \ell'_2 m'}\!+\!F_{\ell_1 \ell_2-m}F_{\ell'_1 \ell'_2 -m'} \!\big\rangle \, .
	\end{eqnarray}
We now evaluate the Gaussian zeroth-order contribution, for which
   \begin{eqnarray}
   F_{\ell_1\ell_2m}=a^{E(P)*}_{\ell_1m}a^{E(P)}_{\ell_2m} \;\;\;\; \textrm{and} \;\;\;\;
   F_{\ell_1\ell_2-m}=a^{E(P)*}_{\ell_1-m}a^{E(P)}_{\ell_2-m} \, .
   \end{eqnarray}
   The first contribution to the four-point function is
	\begin{equation}\label{zerothdefb}
   \begin{split}
   \big\langle f_{\ell_1 \ell_2 m} f_{\ell_1' \ell_2' m'}\big\rangle_{0th} =&\frac{1}{4}\big\langle a^{E*}_{\ell_1m}a^{E}_{\ell_2m}a^{E*}_{\ell'_1m'}a^{E}_{\ell'_2m'}\big\rangle+\frac{1}{4}\big\langle a^{E*}_{\ell_1m}a^{E}_{\ell_2m}a^{E*}_{\ell'_1-m'}a^{E}_{\ell'_2-m'}\big\rangle \,+\\&\frac{1}{4}\big\langle a^{E*}_{\ell_1-m}a^{E}_{\ell_2-m}a^{E*}_{\ell'_1m'}a^{E}_{\ell'_2m'}\big\rangle+\frac{1}{4}\big\langle a^{E*}_{\ell_1-m}a^{E}_{\ell_2-m}a^{E*}_{\ell'_1-m'}a^{E}_{\ell'_2-m'}\big\rangle \, .
   \end{split}
   \end{equation}
For compactness we drop the superscript $(P)$ in what follows. Since the primary modes are Gaussian, Wick's theorem gives
   \begin{eqnarray}\label{Gau_def}
   E[X_1X_2X_3X_4]=E[X_1X_2]E[X_3X_4]+E[X_1X_3]E[X_2X_4]+E[X_1X_4]E[X_2X_3] \,.
   \end{eqnarray} 
  Using Eq.~\eqref{Gau_def}, each term in Eq.~\eqref{zerothdefb} can be decomposed into products of two-point functions. For the first term,
    \begin{equation}
   \begin{split}
   \frac{1}{4}\big\langle a^{E*}_{\ell_1m}a^{E}_{\ell_2m}a^{E*}_{\ell'_1m'}a^{E}_{\ell'_2m'}\big\rangle=& \frac{1}{4}\big\langle a^{E*}_{\ell_1m}a^{E}_{\ell_2m}\big\rangle\big\langle a^{E*}_{\ell'_1m'}a^{E}_{\ell'_2m'}\big\rangle+\frac{1}{4}\big\langle a^{E*}_{\ell_1m}a^{E*}_{\ell'_1m'}\big\rangle\big\langle a^{E}_{\ell_2m}a^{E}_{\ell'_2m'}\big\rangle\\&+\frac{1}{4}\big\langle a^{E*}_{\ell_1m}a^{E}_{\ell'_2m'}\big\rangle\big\langle a^{E}_{\ell_2m}a^{E*}_{\ell'_1m'}\big\rangle \, ,
   \end{split}
   \end{equation}
   which gives
    \begin{equation}
    \begin{split}
  \frac{1}{4}\big\langle a^{E*}_{\ell_1m}a^{E}_{\ell_2m}a^{E*}_{\ell'_1m'}a^{E}_{\ell'_2m'}\big\rangle= &\frac{1}{4}\delta_{\ell_1\ell_2}\delta_{\ell'_1\ell'_2}C^{EE}_{\ell_1}C^{EE}_{\ell'_2}+\frac{1}{4}\delta_{\ell'_2\ell_2}\delta_{\ell_1\ell'_1}\delta_{m-m'}C^{EE}_{\ell_1}C^{EE}_{\ell_2}+\\& \frac{1}{4}\delta_{\ell_1\ell'_2}\delta_{\ell_2\ell'_1}\delta_{mm'}C^{EE}_{\ell_1}C^{EE}_{\ell_2}\;,
  \end{split}
   \end{equation}
The remaining three terms in Eq.~\eqref{zerothdefb} give, respectively,
 \begin{equation}
    \begin{split}
  \frac{1}{4}\big\langle a^{E*}_{\ell_1m}a^{E}_{\ell_2m}a^{E*}_{\ell'_1-m'}a^{E}_{\ell'_2-m'}\big\rangle= &\frac{1}{4}\delta_{\ell_1\ell_2}\delta_{\ell'_1\ell'_2}C^{EE}_{\ell_1}C^{EE}_{\ell'_2}+\frac{1}{4}\delta_{\ell'_2l_2}\delta_{\ell_1\ell'_1}\delta_{mm'}C^{EE}_{\ell_1}C^{EE}_{\ell_2}+\\& \frac{1}{4}\delta_{\ell_1\ell'_2}\delta_{\ell_2\ell'_1}\delta_{m-m'}C^{EE}_{\ell_1}C^{EE}_{\ell_2},
  \end{split}
   \end{equation}

\begin{equation}
    \begin{split}
 \frac{1}{4}\big\langle a^{E*}_{\ell_1-m}a^{E}_{\ell_2-m}a^{E*}_{\ell'_1m'}a^{E}_{\ell'_2m'}\big\rangle= &\frac{1}{4}\delta_{\ell_1\ell_2}\delta_{\ell'_1\ell'_2}C^{EE}_{\ell_1}C^{EE}_{\ell'_2}+\frac{1}{4}\delta_{\ell'_2\ell_2}\delta_{\ell_1\ell'_1}\delta_{mm'}C^{EE}_{\ell_1}C^{EE}_{\ell_2}+\\& \frac{1}{4}\delta_{\ell_1\ell'_2}\delta_{\ell_2\ell'_1}\delta_{-mm'}C^{EE}_{\ell_1}C^{EE}_{\ell_2},
  \end{split}
   \end{equation}

 \begin{equation}
    \begin{split}
  \frac{1}{4}\big\langle a^{E*}_{\ell_1-m}a^{E}_{\ell_2-m}a^{E*}_{\ell'_1-m'}a^{E}_{\ell'_2-m'}\big\rangle= &\frac{1}{4}\delta_{\ell_1\ell_2}\delta_{\ell'_1\ell'_2}C^{EE}_{\ell_1}C^{EE}_{\ell'_2}+\frac{1}{4}\delta_{\ell'_2\ell_2}\delta_{\ell_1\ell'_1}\delta_{m-m'}C^{EE}_{\ell_1}C^{EE}_{\ell_2}+\\& \frac{1}{4}\delta_{\ell_1\ell'_2}\delta_{\ell_2\ell'_1}\delta_{m'm}C^{EE}_{\ell_1}C^{EE}_{\ell_2}.
  \end{split}
   \end{equation}   
Imposing the ordering used in the estimator,
\begin{eqnarray}\label{sym_con}
  \ell_2 > \ell_1, \, \ell'_2 > \ell'_1 \hspace{0.3cm} \text{and} \hspace{0.3cm}\, m'=m, 
\end{eqnarray}
the variance becomes
\begin{equation}\label{zerothdef1}
   \begin{split}
   & \big\langle f_{\ell_1 \ell_2 m} f_{\ell_1' \ell_2' m'}\big\rangle_{0th}-\big\langle f_{\ell_1 \ell_2 m}\big\rangle\big\langle f_{\ell_1' \ell_2' m'}\big\rangle_{0th}
=\frac{1}{2}\delta_{\ell_2\ell'_2}\delta_{\ell_1\ell'_1}C^{EE}_{\ell_1}C^{EE}_{\ell_2}+\frac{1}{2}\delta_{\ell_2\ell'_2}\delta_{\ell_1\ell'_1}\delta_{m-m'}C^{EE}_{\ell_1}C^{EE}_{\ell_2}, 
   \end{split}
   \end{equation}
The disconnected mean contribution is proportional to $\delta_{\ell_1\ell_2}\delta_{\ell'_1\ell'_2}$ and therefore vanishes for the off-diagonal configurations used here. The variance is therefore
\begin{equation}\label{zerothdeff}
   \begin{split}
   \sigma^2_{\ell_1\ell_2m}= \frac{1}{2}\delta_{\ell_2\ell_2'}\delta_{\ell_1\ell_1'}(\delta_{m0}+1)C^{EE}_{\ell_1}C^{EE}_{\ell_2}.
   \end{split}
   \end{equation}
   
\section{The $f^{TE}_{\ell_1\ell_2m}$ component}\label{FTE_complete}
We denote the CMB temperature anisotropy by $T(\hat n)$, with harmonic expansion
\begin{equation}
T(\hat n)=\sum_{\ell m} a^{T}_{\ell m}\,Y_{\ell m}(\hat n).
\end{equation}
For the lensed temperature field,
 \begin{eqnarray}
     \tilde{a}^{\rm T}_{\ell m}=a^{\rm T}_{\ell m}+\sum_{\ell_1 m_1}\sum_{\ell_2 m_2}b_{\ell_1 m_1}a^{\rm T}_{\ell_2 m_2}I_{\ell \ell_1 \ell_2}^{m m_1 m_2} \, ,
 \end{eqnarray}
 where the spin-0 geometrical factor is~\cite{Hu:2000ee}
 \begin{equation}
 \begin{split}
 I^{m m_1 m_2}_{\ell \ell_1 \ell_2}=&\frac{1}{2}[\ell_1(\ell_1+1)+\ell_2(\ell_2+1)-\ell(\ell+1)] \sqrt{\frac{(2\ell_1+1)(2\ell_2+1)(2\ell+1)}{4 \pi}}\times\\&
\begin{pmatrix}
\ell_1 & \ell_2 & \ell_3\\    
0 & 0 & 0
\end{pmatrix}
\begin{pmatrix}
\ell_1 & \ell_2 & \ell_3\\
-m_1 & m_2 & m
\end{pmatrix}
(-1)^{m_1}.
\end{split}
\end{equation}
The polarization field is given by Eq.~\eqref{X_harmonics}.

The $ET$ component can then be written as
\begin{equation}\label{ET}
\begin{split}
    \big\langle a^{\rm E}_{\ell m}\tilde{a}_{\ell^\prime m^\prime}^{\rm T*}\big\rangle=&\frac{1}{2}{}_2\tilde{X}_{\ell m}\tilde{a}_{\ell^\prime m^\prime}^{\rm T*}+ \frac{1}{2}{}_{-2}\tilde{X}_{\ell m}\tilde{a}_{\ell^\prime m^\prime}^{\rm T*} \,.
    \end{split}
\end{equation}
Expanding each factor to first order in the lensing potential gives
\begin{equation}
\begin{split}
    {}_2\tilde{X}_{\ell m}\tilde{a}_{\ell^\prime m^\prime}^{\rm T*}=&\sum_{\ell_1 m_1}\sum_{\ell_2 m_2}b_{\ell_1 m_1}\big\langle{}_2X_{\ell_2 m_2}a_{\ell^\prime m^\prime}^{\rm T*}\big\rangle {}_2I_{\ell \ell_1\ell_2}^{m m_1 m_2}+\\& \sum_{\ell_1 m_1}\sum_{\ell_2 m_2}b_{\ell_1 m_1}\big\langle a_{\ell_2 m_2}^{\rm T*}{}_2X_{\ell m}\big\rangle I_{\ell^\prime \ell_1\ell_2}^{m^\prime m_1 m_2} \, ,
\end{split}
\end{equation}
and 
\begin{equation}
\begin{split}
    {}_{-2}\tilde{X}_{\ell m}\tilde{a}_{\ell^\prime m^\prime}^{\rm T*}=&\sum_{\ell_1 m_1}\sum_{\ell_2 m_2}b_{\ell_1 m_1}\big\langle{}_{-2}X_{\ell_2 m_2}a_{\ell^\prime m^\prime}^{\rm T*}\big\rangle {}_2I_{\ell \ell_1\ell_2}^{m m_1 m_2}+\\&\sum_{\ell_1 m_1}\sum_{\ell_2 m_2}b_{\ell_1 m_1}\big\langle a_{\ell_2 m_2}^{\rm T*}{}_{-2}X_{\ell m}\big\rangle I_{\ell^\prime \ell_1\ell_2}^{m^\prime m_1 m_2} .
\end{split}
\end{equation}
Equivalently, these terms represent the two first-order contractions
\begin{eqnarray}
    {\pm}_2\tilde{X}_{\ell m}\tilde{a}_{\ell^\prime m^\prime}^{\rm T*}=\big\langle a^{E(L_1)*}_{\ell^\prime m^\prime}a^{T}_{\ell m }\big\rangle+\big\langle a^{E}_{\ell m}a^{T (L_1)*}_{\ell^\prime m^\prime }\big\rangle \, .
\end{eqnarray}
Using Eq.~\eqref{X_E_B}, we obtain
\begin{equation}
\begin{split}
   {}_2\tilde{X}_{\ell m}\tilde{a}_{\ell^\prime m^\prime}^{\rm T*}=&\sum_{\ell_1 m_1}\sum_{\ell_2 m_2}b_{\ell_1 m_1}\big\langle a^{\rm E}_{\ell_2 m_2}a_{\ell^\prime m^\prime}^{\rm T*}\big\rangle {}_2I_{\ell \ell_1\ell_2}^{m m_1 m_2}+i\sum_{\ell_1 m_1}\sum_{\ell_2 m_2}b_{\ell_1 m_1}\big\langle a^{\rm B}_{\ell_2 m_2} a_{\ell^\prime m^\prime}^{\rm T*}\big\rangle {}_2I_{\ell \ell_1\ell_2}^{m m_1 m_2} \\& +\sum_{\ell_1 m_1}\sum_{\ell_2 m_2}b_{\ell_1 m_1}\big\langle a^{\rm E}_{\ell m}a_{\ell_2 m_2}^{\rm T *}\big\rangle I_{\ell^\prime \ell_1\ell_2}^{m^\prime m_1 m_2}+i\sum_{\ell_1 m_1}\sum_{\ell_2 m_2}b_{\ell_1 m_1}\big\langle a^{\rm B}_{\ell m} a_{\ell_2 m_2}^{\rm T *}\big\rangle I_{\ell^\prime \ell_1\ell_2}^{m^\prime m_1 m_2} \, ,
\end{split}
\end{equation}
and
\begin{equation}
\begin{split}
   {}_{-2}\tilde{X}_{\ell m}\tilde{a}_{\ell^\prime m^\prime}^{\rm T*}\!=&\sum_{\ell_1 m_1}\sum_{\ell_2 m_2}b_{\ell_1 m_1}\big\langle a^{\rm E}_{\ell_2 m_2}a_{\ell^\prime m^\prime}^{\rm T*}\big\rangle {}_{-2}I_{\ell \ell_1\ell_2}^{m m_1 m_2}\!-\!i\!\sum_{\ell_1 m_1}\sum_{\ell_2 m_2}b_{\ell_1 m_1}\big\langle a^{\rm B}_{\ell_2 m_2} a_{\ell^\prime m^\prime}^{\rm T*}\big\rangle {}_{-2}I_{\ell \ell_1\ell_2}^{m m_1 m_2} \\& \!+\sum_{\ell_1 m_1}\sum_{\ell_2 m_2}b_{\ell_1 m_1}\big\langle a^{\rm E}_{\ell m}a_{\ell_2 m_2}^{\rm T *}\big\rangle I_{\ell^\prime \ell_1\ell_2}^{m^\prime m_1 m_2}\!-\!i\!\sum_{\ell_1 m_1}\sum_{\ell_2 m_2}b_{\ell_1 m_1}\big\langle a^{\rm B}_{\ell m} a_{\ell_2 m_2}^{\rm T *}\big\rangle I_{\ell^\prime \ell_1\ell_2}^{m^\prime m_1 m_2} \, .
\end{split}
\end{equation}
Substituting these expressions into Eq.~\eqref{ET} gives
\begin{equation}
\begin{split}
    \big\langle a^{\rm E}_{\ell m}\tilde{a}_{\ell^\prime m^\prime}^{\rm T*}\big\rangle=&\:\frac{1}{2}\sum_{\ell_1 m_1}\sum_{\ell_2 m_2}b_{\ell_1 m_1}\big\langle a^{\rm E}_{\ell_2 m_2}a_{\ell^\prime m^\prime}^{\rm T*}\big\rangle {}_2I_{\ell \ell_1\ell_2}^{m m_1 m_2}+\frac{1}{2}\sum_{\ell_1 m_1}\sum_{\ell_2 m_2}b_{\ell_1 m_1}\big\langle a^{\rm E}_{\ell_2 m_2}a_{\ell^\prime m^\prime}^{\rm T*}\big\rangle {}_{-2}I_{\ell \ell_1\ell_2}^{m m_1 m_2}\\&+\sum_{\ell_1 m_1}\sum_{\ell_2 m_2}b_{\ell_1 m_1}\big\langle a^{\rm E}_{\ell m}a_{\ell_2 m_2}^{\rm T *}\big\rangle I_{\ell^\prime \ell_1\ell_2}^{m^\prime m_1 m_2}+\frac{1}{2}i\sum_{\ell_1 m_1}\sum_{\ell_2 m_2}b_{\ell_1 m_1}\big\langle a^{\rm B}_{\ell_2 m_2} a_{\ell^\prime m^\prime}^{\rm T*}\big\rangle {}_2I_{\ell \ell_1\ell_2}^{m m_1 m_2}\\&-\frac{1}{2}i\sum_{\ell_1 m_1}\sum_{\ell_2 m_2}b_{\ell_1 m_1}\big\langle a^{\rm B}_{\ell_2 m_2} a_{\ell^\prime m^\prime}^{\rm T*}\big\rangle {}_{-2}I_{\ell \ell_1\ell_2}^{m m_1 m_2} \, .
\end{split}
\end{equation}
Using the relation in Eq.~\eqref{Ipm2}, this becomes
\begin{equation}
\begin{split}
    \big\langle a^{\rm E}_{\ell m}\tilde{a}_{\ell^\prime m^\prime}^{\rm T*}\big\rangle
    =&\:\frac{1}{2}\sum_{\ell_1 m_1}\sum_{\ell_2 m_2}
    b_{\ell_1 m_1}
    \big\langle a^{\rm E}_{\ell_2 m_2}a_{\ell^\prime m^\prime}^{\rm T*}\big\rangle
    {}_2I_{\ell \ell_1\ell_2}^{m m_1 m_2}
    \big(1+(-1)^L\big)
    \\
    &+\sum_{\ell_1 m_1}\sum_{\ell_2 m_2}
    b_{\ell_1 m_1}
    \big\langle a^{\rm E}_{\ell m}a_{\ell_2 m_2}^{\rm T*}\big\rangle
    I_{\ell^\prime \ell_1\ell_2}^{m^\prime m_1 m_2}
    \\
    &+\frac{i}{2}\sum_{\ell_1 m_1}\sum_{\ell_2 m_2}
    b_{\ell_1 m_1}
    \big\langle a^{\rm B}_{\ell_2 m_2}a_{\ell^\prime m^\prime}^{\rm T*}\big\rangle
    {}_{2}I_{\ell \ell_1\ell_2}^{m m_1 m_2}
    \big(1-(-1)^L\big) \, .
\end{split}
\end{equation}
Applying the diagonal two-point functions then gives
\begin{equation}
\begin{split}
   \big\langle a^{\rm E}_{\ell m}\tilde{a}_{\ell^\prime m^\prime}^{\rm T*}\big\rangle=&\:\frac{1}{2}\sum_{\ell_1 m_1}\sum_{\ell_2 m_2}b_{\ell_1 m_1} C^{ET}_{\ell^\prime} {}_2I_{\ell \ell_1\ell_2}^{m m_1 m_2}\big(1+(-1)^L\big)\delta_{\ell_2 \ell^\prime}\delta_{m_2 m^\prime}\\&+\sum_{\ell_1 m_1}\sum_{\ell_2 m_2}b_{\ell_1 m_1} C^{ET}_{\ell} I_{\ell^\prime \ell_1\ell_2}^{m^\prime m_1 m_2}\delta_{\ell \ell_2}\delta_{m m_2}\\&+\frac{1}{2}i\sum_{\ell_1 m_1}\sum_{\ell_2 m_2}b_{\ell_1 m_1} C^{BT}_{\ell^\prime} {}_{2}I_{\ell \ell_1\ell_2}^{m m_1 m_2}\big(1-(-1)^L\big)\delta_{\ell^\prime \ell_2}\delta_{m^\prime m_2} \, .
\end{split}
\end{equation}
Combining these terms and aligning the $\hat z$ axis with the center of the lensing profile, so that only $m_1=0$ contributes, gives
\begin{equation}
\begin{split}
   \big\langle a^{\rm E}_{\ell m}\tilde{a}_{\ell^\prime m^\prime}^{\rm T*}\big\rangle=&\:\frac{1}{2}C^{ET}_{\ell^\prime}\sum_{\ell_1}b_{\ell_1 0} \big(1+(-1)^L\big) {}_2I_{\ell \ell_1\ell^\prime}^{m 0 m^\prime}+C^{ET}_{\ell}\sum_{\ell_1}b_{\ell_1 0}\hspace{0.15cm} I_{\ell^\prime \ell_1\ell}^{m^\prime 0 m}\\&+\frac{i}{2}C^{BT}_{\ell^\prime}\sum_{\ell_1}b_{\ell_1 0}\hspace{0.15cm} {}_{2}I_{\ell \ell_1\ell^\prime}^{m 0 m^\prime}\big(1-(-1)^L\big)  \, .
\end{split}
\end{equation}
Equivalently, this expression can be written as
\begin{equation}
\begin{split}
   \big\langle a^{E}_{\ell m}a^{T (L_1)*}_{\ell^\prime m^\prime }\big\rangle+\big\langle a^{T(L_1)*}_{\ell^\prime m^\prime}a^{E}_{\ell m }\big\rangle=&\:\frac{1}{2}C^{ET}_{\ell^\prime}\sum_{\ell_1}b_{\ell_1 0} \big(1+(-1)^L\big) {}_2I_{\ell \ell_1\ell^\prime}^{m 0 m^\prime}+C^{ET}_{\ell}\sum_{\ell_1}b_{\ell_1 0}\hspace{0.15cm} I_{\ell^\prime \ell_1\ell}^{m^\prime 0 m}\\&+\frac{i}{2}C^{BT}_{\ell^\prime}\sum_{\ell_1}b_{\ell_1 0}\hspace{0.15cm} {}_{2}I_{\ell \ell_1\ell^\prime}^{m 0 m^\prime}\big(1-(-1)^L\big) \, ,
\end{split}
\end{equation}
where $a^{(L_1)}_{\ell m}$ denotes the contribution linear in the lensing potential. In the parity-even, negligible-primordial-$B$ limit the imaginary $BT$ term is dropped, leaving
\begin{equation}
\begin{split}
   F^{TH(ET)}_{\ell \ell^\prime m}=&\frac{1}{2}C^{ET}_{\ell^\prime}\sum_{\ell_1}b_{\ell_1 0} \big(1+(-1)^L\big) {}_2I_{\ell \ell_1\ell^\prime}^{m 0 m^\prime}+C^{ET}_{\ell}\sum_{\ell_1}b_{\ell_1 0}\hspace{0.15cm} I_{\ell^\prime \ell_1\ell}^{m^\prime 0 m} \, .
\end{split}
\end{equation}
The final ordered $ET$ response can therefore be expressed as
\begin{equation}
\begin{split}
   F^{TH(ET)}_{\ell_1 \ell_2 m} &=\big\langle a^{E}_{\ell_1m_1}a^{T(L_1)*}_{\ell_2m_2}\big\rangle + \big\langle a^{T*}_{\ell_1m_1}a^{E(L_1)}_{\ell_2 m_2}\big\rangle \\&=\frac{(-1)^m}{2}C^{ET}_{\ell_2}\sum_{\ell_3}{}^2G_{\ell_1\ell_2\ell_3}^{-m m 0} \big(1+(-1)^L\big) \frac{\ell_2(\ell_2+1)-\ell_1(\ell_1+1)+\ell_3(\ell_3+1)}{2}b_{\ell_3 0}\\&+(-1)^m C^{ET}_{\ell_1}\sum_{\ell_3}\hspace{0.15cm} {}^0G_{\ell_1\ell_2\ell_3}^{-m m 0} \frac{\ell_1(\ell_1+1)-\ell_2(\ell_2+1)+\ell_3(\ell_3+1)}{2}b_{\ell_3 0} \,,
\end{split}
\end{equation}
where
\begin{equation}
\begin{split}
{}^0G_{\ell_1\ell_2\ell_3}^{-mm0}=\sqrt{\frac{(2\ell_1+1)(2\ell_2+1)(2\ell_3+1)}{4\pi}}
\begin{pmatrix}
\ell_1 & \ell_2 & \ell_3\\    
0 & 0 & 0
\end{pmatrix}
\begin{pmatrix}
\ell_1 & \ell_2 & \ell_3\\
-m & m & 0
\end{pmatrix}
,
\end{split}
\end{equation}

\section{Variance of $f^{TE}_{\ell_1\ell_2m}$}\label{VarTE}
For the variance of the crossed $TE$ pair, we introduce
\begin{eqnarray}
    F^{TE}_{\ell_1\ell_2m} \equiv a^{T*}_{\ell_1m}a^{E}_{\ell_2m} \, ,
\end{eqnarray}
\begin{eqnarray}
    F^{TE}_{\ell_1\ell_2-m}=a^{T*}_{\ell_1-m}a^{E}_{\ell_2-m} \,,
\end{eqnarray}
with
\begin{eqnarray}
a^T_{\ell m} = a_{\ell m}^{T(P)}+a_{\ell m}^{T(L)} \, , 
\end{eqnarray}
and 
\begin{eqnarray}
a^{\rm E}_{\ell m} = a_{\ell m}^{E(P)}+a_{\ell m}^{E(L)} .
\end{eqnarray}

 Using the same real-mode decomposition, the zeroth-order Gaussian contribution is built from
   \begin{eqnarray}
   F_{\ell_1\ell_2m}=a^{T(P)*}_{\ell_1m}a^{E(P)}_{\ell_2m} \, ,
   \end{eqnarray}
    \begin{eqnarray}
   F_{\ell_1\ell_2-m}=a^{T(P)*}_{\ell_1-m}a^{E(P)}_{\ell_2-m} \, ,
   \end{eqnarray}
	\begin{equation}\label{zerothdef}
   \begin{split}
   \big\langle f_{\ell_1 \ell_2 m} f_{\ell_1' \ell_2' m'}\big\rangle_{0th} &=\frac{1}{4}\big\langle a^{T*}_{\ell_1m}a^{E}_{\ell_2m}a^{T*}_{\ell'_1m'}a^{E}_{\ell'_2m'}\big\rangle+\frac{1}{4}\big\langle a^{T*}_{\ell_1m}a^{E}_{\ell_2m}a^{T*}_{\ell'_1-m'}a^{E}_{\ell'_2-m'}\big\rangle\\&+\frac{1}{4}\big\langle a^{T*}_{\ell_1-m}a^{E}_{\ell_2-m}a^{T*}_{\ell'_1m'}a^{E}_{\ell'_2m'}\big\rangle+\frac{1}{4}\big\langle a^{T*}_{\ell_1-m}a^{E}_{\ell_2-m}a^{T*}_{\ell'_1-m'}a^{E}_{\ell'_2-m'}\big\rangle \,.
   \end{split}
   \end{equation}
Using Wick's theorem, Eq.~\eqref{Gau_def}, each term in Eq.~\eqref{zerothdef} decomposes into products of two-point functions. For the first term,
    \begin{equation}
   \begin{split}
   \frac{1}{4}\big\langle a^{T*}_{\ell_1m}a^{E}_{\ell_2m}a^{T*}_{\ell'_1m'}a^{E}_{\ell'_2m'}\big\rangle=& \frac{1}{4}\big\langle a^{T*}_{\ell_1m}a^{E}_{\ell_2m}\big\rangle\big\langle a^{T*}_{\ell'_1m'}a^{E}_{\ell'_2m'}\big\rangle+\frac{1}{4}\big\langle a^{T*}_{\ell_1m}a^{T*}_{\ell'_1m'}\big\rangle\big\langle a^{E}_{\ell_2m}a^{E}_{\ell'_2m'}\big\rangle\\&+\frac{1}{4}\big\langle a^{T*}_{\ell_1m}a^{E}_{\ell'_2m'}\big\rangle\big\langle a^{E}_{\ell_2m}a^{T*}_{\ell'_1m'}\big\rangle \,,
   \end{split}
   \end{equation}
   which gives
    \begin{equation}
    \begin{split}
  \frac{1}{4}\big\langle a^{T*}_{\ell_1m}a^{E}_{\ell_2m}a^{T*}_{\ell'_1m'}a^{E}_{\ell'_2m'}\big\rangle= &\frac{1}{4}\delta_{\ell_1\ell_2}\delta_{\ell'_1\ell'_2}C^{TE}_{\ell_1}C^{TE}_{\ell'_2}+\frac{1}{4}\delta_{\ell'_2\ell_2}\delta_{\ell_1\ell'_1}\delta_{m-m'}C^{TT}_{\ell_1}C^{EE}_{\ell_2}+\\& \frac{1}{4}\delta_{\ell_1\ell'_2}\delta_{\ell_2\ell'_1}\delta_{mm'}C^{TE}_{\ell_1}C^{TE}_{\ell_2},
  \end{split}
   \end{equation}
The remaining three terms in Eq.~\eqref{zerothdef} give, respectively,
 \begin{equation}
    \begin{split}
  \frac{1}{4}\big\langle a^{T*}_{\ell_1m}a^{E}_{\ell_2m}a^{T*}_{\ell'_1-m'}a^{E}_{\ell'_2-m'}\big\rangle= &\frac{1}{4}\delta_{\ell_1\ell_2}\delta_{\ell'_1\ell'_2}C^{TE}_{\ell_1}C^{TE}_{\ell'_2}+\frac{1}{4}\delta_{\ell'_2\ell_2}\delta_{\ell_1\ell'_1}\delta_{mm'}C^{TT}_{\ell_1}C^{EE}_{\ell_2}+\\& \frac{1}{4}\delta_{\ell_1\ell'_2}\delta_{\ell_2\ell'_1}\delta_{m-m'}C^{TE}_{\ell_1}C^{TE}_{\ell_2},
  \end{split}
   \end{equation}

\begin{equation}
    \begin{split}
 \frac{1}{4}\big\langle a^{T*}_{\ell_1-m}a^{E}_{\ell_2-m}a^{T*}_{\ell'_1m'}a^{E}_{\ell'_2m'}\big\rangle= &\frac{1}{4}\delta_{\ell_1\ell_2}\delta_{\ell'_1\ell'_2}C^{TE}_{\ell_1}C^{TE}_{\ell'_2}+\frac{1}{4}\delta_{\ell'_2\ell_2}\delta_{\ell_1\ell'_1}\delta_{mm'}C^{TT}_{\ell_1}C^{EE}_{\ell_2}+\\& \frac{1}{4}\delta_{\ell_1\ell'_2}\delta_{\ell_2\ell'_1}\delta_{-mm'}C^{TE}_{\ell_1}C^{TE}_{\ell_2},
  \end{split}
   \end{equation}

 \begin{equation}
    \begin{split}
  \frac{1}{4}\big\langle a^{T*}_{\ell_1-m}a^{E}_{\ell_2-m}a^{T*}_{\ell'_1-m'}a^{E}_{\ell'_2-m'}\big\rangle= &\frac{1}{4}\delta_{\ell_1\ell_2}\delta_{\ell'_1\ell'_2}C^{TE}_{\ell_1}C^{TE}_{\ell'_2}+\frac{1}{4}\delta_{\ell'_2\ell_2}\delta_{\ell_1\ell'_1}\delta_{m-m'}C^{TT}_{\ell_1}C^{EE}_{\ell_2}+\\& \frac{1}{4}\delta_{\ell_1\ell'_2}\delta_{\ell_2\ell'_1}\delta_{-m'-m}C^{TE}_{\ell_1}C^{TE}_{\ell_2}.
  \end{split}
   \end{equation}   
Using the ordering in Eq.~\eqref{sym_con}, the disconnected mean term again vanishes for the off-diagonal configurations. The crossed-pair variance is therefore
\begin{equation}\label{zerothdeff_TE}
   \begin{split}
   \sigma^2_{\ell_1\ell_2m}= \frac{1}{2}\delta_{\ell_2\ell'_2}\delta_{\ell_1\ell'_1}(\delta_{m0}+1)C^{TT}_{\ell_1}C^{EE}_{\ell_2}.
   \end{split}
   \end{equation}

\section{The $f^{TB}_{\ell_1\ell_2 m}$ component}\label{FTB_complete}

We follow the same steps as in Appendix~\ref{FTE_complete}, now for the $TB$ channel.

The lensed temperature harmonic coefficients are
\begin{equation}
\tilde{a}^{\rm T}_{\ell m}=a^{\rm T}_{\ell m}+\sum_{\ell_1 m_1}\sum_{\ell_2 m_2}b_{\ell_1 m_1}a^{\rm T}_{\ell_2 m_2}I_{\ell \ell_1 \ell_2}^{m m_1 m_2}.
\end{equation}
For polarization, the lensed spin-$\pm 2$ fields satisfy
\begin{equation}\label{spin_lensed}
{}_{\pm2}\tilde{X}_{\ell m}={}_{\pm2}X_{\ell m}+\sum_{\ell_1 m_1}\sum_{\ell_2 m_2}b_{\ell_1 m_1}\,{}_{\pm2}X_{\ell_2 m_2}\,{}_{\pm2}I_{\ell \ell_1 \ell_2}^{m m_1 m_2}.
\end{equation}
We define the lensed $E$ and $B$ modes as
\begin{equation}\label{EB_def_spin}
\tilde{a}^{\rm E}_{\ell m}=\frac12\left({}_{2}\tilde{X}_{\ell m}+{}_{-2}\tilde{X}_{\ell m}\right),\qquad
\tilde{a}^{B}_{\ell m}=\frac{1}{2i}\left({}_{2}\tilde{X}_{\ell m}-{}_{-2}\tilde{X}_{\ell m}\right),
\end{equation}
and similarly for the unlensed fields. We will use Eq.~\eqref{Ipm2} in the form
\begin{equation}\label{Ipm2_used}
{}_{-2}I_{\ell \ell_1 \ell_2}^{m m_1 m_2}=(-1)^{L}\,{}_{2}I_{\ell \ell_1 \ell_2}^{m m_1 m_2},
\qquad L\equiv \ell+\ell_1+\ell_2.
\end{equation}

The $TB$ correlation can be written as
\begin{equation}
\begin{split}
\left\langle \tilde{a}^{B}_{\ell m}\tilde{a}_{\ell' m'}^{\rm T *}\right\rangle
&=\frac{1}{2i}\left\langle {}_{2}\tilde{X}_{\ell m}\tilde{a}_{\ell' m'}^{\rm T *}\right\rangle
-\frac{1}{2i}\left\langle {}_{-2}\tilde{X}_{\ell m}\tilde{a}_{\ell' m'}^{\rm T *}\right\rangle.
\end{split}
\end{equation}
Keeping only terms linear in $b_{\ell_1 m_1}$, each contribution is
\begin{equation}
\begin{split}
{}_{\pm2}\tilde{X}_{\ell m}\tilde{a}_{\ell' m'}^{\rm T *}
=&\sum_{\ell_1 m_1}\sum_{\ell_2 m_2}b_{\ell_1 m_1}\left\langle {}_{\pm2}X_{\ell_2 m_2}a_{\ell' m'}^{\rm T *}\right\rangle {}_{\pm2}I_{\ell \ell_1\ell_2}^{m m_1 m_2}\\
&+\sum_{\ell_1 m_1}\sum_{\ell_2 m_2}b_{\ell_1 m_1}\left\langle a^{\rm T *}_{\ell_2 m_2}\,{}_{\pm2}X_{\ell m}\right\rangle I_{\ell' \ell_1\ell_2}^{m' m_1 m_2}.
\end{split}
\end{equation}
Using ${}_{\pm2}X_{\ell m}=a^{\rm E}_{\ell m}\pm ia^{\rm B}_{\ell m}$, the diagonal correlators give
\begin{equation}
\left\langle {}_{\pm2}X_{\ell_2 m_2}a_{\ell' m'}^{\rm T *}\right\rangle
=\delta_{\ell_2\ell'}\delta_{m_2 m'}\left(C_{\ell'}^{ET}\pm i\,C_{\ell'}^{BT}\right),
\end{equation}
and
\begin{equation}
\left\langle a^{\rm T *}_{\ell_2 m_2}\,{}_{\pm2}X_{\ell m}\right\rangle
=\delta_{\ell_2\ell}\delta_{m_2 m}\left(C_{\ell}^{TE}\pm i\,C_{\ell}^{TB}\right).
\end{equation}
Substituting into the expression for $\langle \tilde{B}\tilde{T}^{*}\rangle$ and combining the ${}_{2}$ and ${}_{-2}$ pieces,
the $ET$ terms survive weighted by the difference ${}_{2}I-{}_{-2}I$, while the $TB$ terms survive weighted by the sum ${}_{2}I+{}_{-2}I$.
Using Eq.~\eqref{Ipm2_used}, we obtain
\begin{equation}
\begin{split}
\left\langle \tilde{a}^{B}_{\ell m}\tilde{a}_{\ell' m'}^{\rm T *}\right\rangle
=&\frac{1}{2}\,C_{\ell'}^{ET}\sum_{\ell_1 m_1} b_{\ell_1 m_1}\,{}_{2}I_{\ell \ell_1\ell'}^{m m_1 m'}\big(1-(-1)^L\big)
+ C_{\ell}^{TB}\sum_{\ell_1 m_1} b_{\ell_1 m_1}\, I_{\ell' \ell_1\ell}^{m' m_1 m},
\end{split}
\end{equation}
where, in the first term, $L\equiv \ell+\ell_1+\ell'$.

Aligning the $\hat z$ axis with the center of the lensing profile gives $m_1=0$ and $b_{\ell_1 m_1}\to b_{\ell_1 0}$.
Dropping the parity-odd spectrum $C_\ell^{TB}$, which vanishes in the parity-even fiducial cosmology, the theoretical $TB$ off-diagonal correlator becomes
\begin{equation}
\left\langle \tilde{a}^{B}_{\ell m}\tilde{a}_{\ell' m'}^{\rm T *}\right\rangle
=\frac{1}{2}C_{\ell'}^{ET}\sum_{\ell_1} b_{\ell_1 0}\,{}_{2}I_{\ell \ell_1\ell'}^{m 0 m'}\big(1-(-1)^L\big).
\end{equation}
Converting ${}_{2}I$ to the Gaunt-integral form, and working in the strict first-order, negligible-primordial-$B$ limit, we obtain
\begin{equation}\label{FTH_TB_final}
\begin{split}
f^{\rm TH(TB)}_{\ell_1\ell_2 m}
=&\frac{(-1)^m}{2}\,C_{\ell_2}^{ET}
\sum_{\ell_3} {}^{2}G_{\ell_1\ell_2\ell_3}^{-m\,m\,0}\,\big(1-(-1)^L\big)\,
\frac{\ell_2(\ell_2+1)-\ell_1(\ell_1+1)+\ell_3(\ell_3+1)}{2}\,b_{\ell_3 0}.
\end{split}
\end{equation}
\section{Variance of $f^{TB}_{\ell_1\ell_2m}$}\label{VarTB}

The variance is defined as $\sigma^2_{\ell_1\ell_2m}=\langle f^2_{\ell_1\ell_2m}\rangle-\langle f_{\ell_1\ell_2m}\rangle^2$. For the $TB$ cross-correlation we define
\begin{eqnarray}
    F^{TB}_{\ell_1\ell_2m} \equiv a^{T*}_{\ell_1m}a^{B}_{\ell_2m},\qquad
    F^{TB}_{\ell_1\ell_2,-m}=a^{T*}_{\ell_1,-m}a^{B}_{\ell_2,-m}.
\end{eqnarray}
Working with real quantities,
\begin{eqnarray}
f_{\ell_1\ell_2m}\equiv \frac{1}{2}\left(F_{\ell_1\ell_2m}+F_{\ell_1\ell_2,-m}\right).
\end{eqnarray}
At zeroth order in the lensing potential, the Gaussian fields satisfy
\begin{eqnarray}
\langle a^X_{\ell m} a^{Y*}_{\ell' m'}\rangle=\delta_{\ell\ell'}\delta_{mm'}\,C^{XY}_\ell.
\end{eqnarray}
Imposing the selection used in the main text ($\ell_2>\ell_1$ and $m'=m$), all terms proportional to $\delta_{\ell_1\ell_2}$ vanish, so that
\begin{equation}
\sigma^2_{\ell_1\ell_2m}(TB)=\frac{1}{2}\,\delta_{\ell_1\ell_1'}\delta_{\ell_2\ell_2'}\left(\delta_{m0}+1\right)\,C^{TT}_{\ell_1}\,C^{BB}_{\ell_2}.
\end{equation}
In parity-even $\Lambda$CDM, $C_\ell^{TB}=0$, so the mean vanishes. The covariance nevertheless contains $C_\ell^{BB}$, which in the forecast includes the lensed $B$-mode power and instrumental noise.

\section{The $f^{EB}_{\ell_1\ell_2 m}$ component}\label{FEB_complete}

We next derive the lensing-induced $EB$ off-diagonal correlator.

Using Eq.~\eqref{EB_def_spin}, the $EB$ correlator can be written as
\begin{equation}
\begin{split}
\left\langle a^{\rm E}_{\ell m}\tilde{B}_{\ell' m'}^{*}\right\rangle
&=\frac{1}{4i}\Big[
\left\langle {}_{2}\tilde{X}_{\ell m}{}_{2}\tilde{X}_{\ell' m'}^{*}\right\rangle
-\left\langle {}_{2}\tilde{X}_{\ell m}{}_{-2}\tilde{X}_{\ell' m'}^{*}\right\rangle\\
&\hspace{1.45cm}
+\left\langle {}_{-2}\tilde{X}_{\ell m}{}_{2}\tilde{X}_{\ell' m'}^{*}\right\rangle
-\left\langle {}_{-2}\tilde{X}_{\ell m}{}_{-2}\tilde{X}_{\ell' m'}^{*}\right\rangle
\Big].
\end{split}
\end{equation}

Expanding to first order in $b_{\ell_1 m_1}$ using Eq.~\eqref{spin_lensed}, each correlator has the form
\begin{equation}
\begin{split}
\left\langle {}_{s}\tilde{X}_{\ell m}\,{}_{s'}\tilde{X}_{\ell' m'}^{*}\right\rangle
=&\sum_{\ell_1 m_1}\sum_{\ell_2 m_2} b_{\ell_1 m_1}\left\langle {}_{s}X_{\ell_2 m_2}\,{}_{s'}X_{\ell' m'}^{*}\right\rangle {}_{s}I_{\ell \ell_1\ell_2}^{m m_1 m_2}\\
&+\sum_{\ell_1 m_1}\sum_{\ell_2 m_2} b_{\ell_1 m_1}\left\langle {}_{s}X_{\ell m}\,{}_{s'}X_{\ell_2 m_2}^{*}\right\rangle {}_{s'}I_{\ell' \ell_1\ell_2}^{m' m_1 m_2}.
\end{split}
\end{equation}
Using ${}_{2}X=E+iB$ and ${}_{-2}X=E-iB$, the diagonal correlators are
\begin{align}
\left\langle {}_{2}X_{\ell m}{}_{2}X_{\ell' m'}^{*}\right\rangle
&=\delta_{\ell\ell'}\delta_{mm'}\left(C_\ell^{EE}+C_\ell^{BB}\right),\\
\left\langle {}_{2}X_{\ell m}{}_{-2}X_{\ell' m'}^{*}\right\rangle
&=\delta_{\ell\ell'}\delta_{mm'}\left(C_\ell^{EE}-C_\ell^{BB}+2i\,C_\ell^{EB}\right),\\
\left\langle {}_{-2}X_{\ell m}{}_{2}X_{\ell' m'}^{*}\right\rangle
&=\delta_{\ell\ell'}\delta_{mm'}\left(C_\ell^{EE}-C_\ell^{BB}-2i\,C_\ell^{EB}\right).
\end{align}
Combining the terms and using Eq.~\eqref{Ipm2_used}, the $EB$ signal is sourced by the difference between the spin-$+2$ and spin-$-2$ coupling kernels, yielding a parity-odd selection factor $(1-(-1)^L)$.

Aligning $\hat z$ with the lens center ($m_1=0$) and rewriting the result in Gaunt-integral form gives
\begin{equation}\label{FTH_EB_final_app}
\begin{split}
f^{\rm TH(EB)}_{\ell_1\ell_2 m}
=&\frac{(-1)^m}{2}
\sum_{\ell_3} {}^{2}G_{\ell_1\ell_2\ell_3}^{-m\,m\,0}\,\big(1-(-1)^L\big)\,
\Bigg[
C_{\ell_1}^{EE}\,\frac{\ell_1(\ell_1+1)-\ell_2(\ell_2+1)+\ell_3(\ell_3+1)}{2}\\
&\hspace{3.55cm}
-\,C_{\ell_2}^{BB}\,\frac{\ell_2(\ell_2+1)-\ell_1(\ell_1+1)+\ell_3(\ell_3+1)}{2}
\Bigg]\,b_{\ell_3 0}.
\end{split}
\end{equation}
In standard parity-even cosmology $C_\ell^{EB}=0$, and for scalar primordial perturbations $C_\ell^{BB}\simeq0$ at leading order. The dominant contribution is therefore the term proportional to $C_{\ell_1}^{EE}$. Lensing-generated $BB$ power arises at higher order in the lensing potential, and in the strict first-order, negligible-primordial-$B$ limit this reduces to

\begin{equation}\label{FTH_EB_1st_order}
\begin{split}
f^{\rm TH(EB)}_{\ell_1\ell_2 m}
=&\frac{(-1)^m}{2}
\sum_{\ell_3} {}^{2}G_{\ell_1\ell_2\ell_3}^{-m\,m\,0}\,\big(1-(-1)^L\big)\,
C_{\ell_1}^{EE}\,\frac{\ell_1(\ell_1+1)-\ell_2(\ell_2+1)+\ell_3(\ell_3+1)}{2}b_{\ell_3 0} \,.
\end{split}
\end{equation}


\section{Variance of $f^{EB}_{\ell_1\ell_2m}$}\label{VarEB}

For the $EB$ cross-correlation we define
\begin{eqnarray}
    F^{EB}_{\ell_1\ell_2m} \equiv a^{E*}_{\ell_1m}a^{B}_{\ell_2m},\qquad
    F^{EB}_{\ell_1\ell_2,-m}=a^{E*}_{\ell_1,-m}a^{B}_{\ell_2,-m},
\end{eqnarray}
and again $f_{\ell_1\ell_2m}\equiv \frac{1}{2}(F_{\ell_1\ell_2m}+F_{\ell_1\ell_2,-m})$. At zeroth order, using the same selection $\ell_2>\ell_1$ and $m'=m$, we obtain
\begin{equation}
\sigma^2_{\ell_1\ell_2m}(EB)=\frac{1}{2}\,\delta_{\ell_1\ell_1'}\delta_{\ell_2\ell_2'}\left(\delta_{m0}+1\right)\,C^{EE}_{\ell_1}\,C^{BB}_{\ell_2}.
\end{equation}

\section{The $f^{BB}_{\ell_1\ell_2 m}$ component}\label{FBB_complete}

Finally, we consider the $BB$ off-diagonal correlator. Proceeding as above, we write
\begin{equation}
\left\langle \tilde{a}^{B}_{\ell m}\tilde{B}_{\ell' m'}^{*}\right\rangle
=\frac{1}{4}\Big[
\left\langle {}_{2}\tilde{X}_{\ell m}{}_{2}\tilde{X}_{\ell' m'}^{*}\right\rangle
-\left\langle {}_{2}\tilde{X}_{\ell m}{}_{-2}\tilde{X}_{\ell' m'}^{*}\right\rangle
-\left\langle {}_{-2}\tilde{X}_{\ell m}{}_{2}\tilde{X}_{\ell' m'}^{*}\right\rangle
+\left\langle {}_{-2}\tilde{X}_{\ell m}{}_{-2}\tilde{X}_{\ell' m'}^{*}\right\rangle
\Big].
\end{equation}
At first order in $b_{\ell_1 m_1}$, the surviving terms are weighted by the sum ${}_{2}I+{}_{-2}I$. Using Eq.~\eqref{Ipm2_used}, this gives the parity-even selector $(1+(-1)^L)$.

After aligning $\hat z$ with the lens center ($m_1=0$) and converting to Gaunt integrals, we obtain
\begin{equation}\label{FTH_BB_final}
\begin{split}
f^{\rm TH(BB)}_{\ell_1\ell_2 m}
=&\frac{(-1)^m}{2}
\sum_{\ell_3} {}^{2}G_{\ell_1\ell_2\ell_3}^{-m\,m\,0}\,\big(1+(-1)^L\big)\,
\Bigg[
C_{\ell_1}^{BB}\,\frac{\ell_1(\ell_1+1)-\ell_2(\ell_2+1)+\ell_3(\ell_3+1)}{2}\\
&\hspace{3.55cm}
+\,C_{\ell_2}^{BB}\,\frac{\ell_2(\ell_2+1)-\ell_1(\ell_1+1)+\ell_3(\ell_3+1)}{2}
\Bigg]\,b_{\ell_3 0} \simeq 0,
\end{split}
\end{equation}
For negligible primordial $B$ modes, $C_\ell^{BB}\simeq0$ at leading order. The first-order $BB$ response therefore vanishes, consistent with the fact that lensing-generated $BB$ power is a higher-order effect in the lensing potential.

\vspace{4mm}   
\noindent\emph{\textbf{Acknowledgments.}}
PSF acknowledges the support from the start-up funding of Zhejiang University, Zhejiang provincial top level research support program, FAPES (Brazil) and FINEP/FACC (contract 01.22.0505.00). The analysis presented in this article was carried out on the SilkRiver Supercomputer of Zhejiang University (China) and the \href{https://computacaocientifica.ufes.br/scicom}{Sci-Com Lab} Supercomputer of the Federal University of Espírito Santo (Brazil, supported by FAPES, CAPES, and CNPq).
The Wigner $3j$ symbols are evaluated with the \textsc{pywigxjpf}/\textsc{WIGXJPF} code~\cite{Johansson:2015cca}.

\bibliographystyle{JHEP2015}
\bibliography{cmb}

\end{document}